%% file: main.tex
\documentclass[a4paper,12pt]{article}

\usepackage[utf8]{inputenc}
\usepackage{amsmath, amsfonts, amssymb, amsbsy, bm} 
\usepackage[
  backend=biber,
  style=apa,
  natbib=true,
  uniquelist=false,
  doi=false,
  url=false,
  isbn=false
]{biblatex}

\DeclareSourcemap{
  \maps[datatype=bibtex]{
    \map{
      \step[fieldset=note, null]
      \step[fieldset=abstract, null]
      \step[fieldset=urldate, null]
      \step[fieldset=url, null]
    }
  }
}

\usepackage{CJKutf8} 
\usepackage{booktabs, dcolumn}  
\usepackage{graphicx, epstopdf} 
\graphicspath{{Figure/}}
\usepackage[inkscapelatex=false]{svg} 
\usepackage{pdflscape}          
\usepackage{longtable}          
\usepackage{appendix}           
\usepackage{dashbox}            
\usepackage{hyperref}           
\usepackage{subfig}             
\usepackage{psfrag}             
\usepackage{float}              
\usepackage{multirow}           
\usepackage{enumerate}          
\usepackage{pgffor}             
\usepackage{tabularx, colortbl} 
\usepackage{tikz}               
\usepackage{cancel}             
\usepackage[normalem]{ulem}     
\usepackage{threeparttable}
\usepackage{arydshln}
\usepackage[table]{xcolor}

\usepackage{caption}
\newcolumntype{z}[1]{D{.}{.}{#1}}

\title{\Large\bfseries Financial Volatility and Risk Forecasting Incorporating a Larger Number of Realized Measures}
\author{
Qianli Zhao\footnote{School of Business Analytics and Marketing, The University of Sydney}\hspace{0.2em}
\thanks{Corresponding author. Email: miles.zhao@sydney.edu.au}\and
Chao Wang\footnotemark[1] \and
Richard Gerlach\footnotemark[1] \and
Giuseppe Storti\footnote{Department of Economics and Statistics, University of Salerno} \and
Lingxiang Zhang\footnote{School of Economics, Beijing Institute of Technology}
}
\date{}
\hypersetup{
  pdftitle={Financial Volatility and Risk Forecasting Incorporating a Larger Number of Realized Measures},
  pdfauthor={},
  pdfsubject={},
  pdfkeywords={}
}

\begin{document}
\maketitle

\begin{abstract}
High-frequency data have led to the development of numerous realized volatility measures, while standard Realized GARCH models typically rely on a single measure. This paper proposes three approaches to incorporate information from multiple realized measures within the Realized GARCH framework. First, we develop an autoencoder-enhanced Realized GARCH model that constructs a synthetic realized measure through nonlinear dimensionality reduction. Second, we propose a linear weighted Realized GARCH model that jointly estimates convex aggregation weights and model parameters. Third, we introduce a functional weighted Realized GARCH model based on parsimonious Beta weighting functions. Using approximately 22 years of data for 29 international equity indices, we evaluate the proposed methods in volatility, Value-at-Risk, and Expected Shortfall forecasting. Compared with benchmark models based on single realized measures, equal weighting, and linear dimension reduction, the proposed methods consistently deliver more accurate volatility and risk forecasts.
\end{abstract}



\section{Introduction}

Volatility forecasting is a key component of financial decision-making, with important applications in risk management, asset pricing, and portfolio allocation. Accurate forecasts provide crucial information for setting capital reserves, valuing derivatives, and controlling market exposure. They are also central to financial risk forecasting because conditional volatility determines the scale of the return distribution used to compute Value-at-Risk (VaR) and Expected Shortfall (ES).

Generalized autoregressive conditional heteroskedasticity (GARCH) captures time-varying volatility by modeling conditional variance as a function of past squared returns and past conditional variance \citep{bollerslev1986}. Since volatility is latent, GARCH models commonly use squared daily returns as an ex post proxy for volatility, although this proxy contains measurement noise and does not use intraday information \citep{andersen1998}.

High-frequency intraday data enable the construction of realized measures, which provide informative proxies for daily volatility. Realized variance (RV) uses intraday returns to estimate daily price variation \citep{andersen2003}. Realized kernels (RK) address market microstructure noise \citep{barndorff-nielsen2009}. Bipower variation (BV) separates continuous variation from jump variation \citep{barndorff-nielsen2004}. Realized semivariance (RSV) captures downside variation \citep{barndorff-nielsen2008b}. Median realized variance (MedRV) improves robustness to jumps and outliers \citep{andersen2012}.

Realized GARCH incorporates realized measures through a measurement equation that links the observed realized measure to latent conditional variance \citep{hansen2012}. This structure jointly models returns and realized volatility, improving on GARCH models that use daily returns alone. However, the standard Realized GARCH framework typically uses only one realized measure at a time. This univariate realized measure structure becomes restrictive when many realized measures are available, because selecting a single measure from a large set of candidates introduces an additional source of model uncertainty. Realized Exponential GARCH (RealEGARCH) addresses this limitation by allowing multiple realized measures to enter the model through additional measurement equations \citep{hansen2016a}. However, as the number of candidate measures increases, adding separate measurement equations becomes less tractable.

A related solution is to combine several realized volatility estimators before using them in the forecasting model. \citet{patton2009} show that combining realized volatility estimators can improve volatility measurement because individual estimators do not fully contain the information in the others. A parsimonious alternative is therefore to construct a synthetic realized measure from several candidate measures. \citet{naimoli2022} use Principal Component Analysis (PCA) and Independent Component Analysis (ICA) to construct synthetic realized measures before incorporating them into the Realized GARCH framework, yielding PC-RealGARCH and IC-RealGARCH, respectively. Both models retain a univariate realized measure input. However, PCA and ICA rely on linear transformations, whereas the relationships among microstructure noise, jumps, persistence, and downside volatility may be nonlinear.

This limitation motivates the use of nonlinear representation learning. Autoencoders are neural networks that learn low-dimensional representations by reconstructing high-dimensional inputs through a bottleneck layer \citep{hinton2006}. In finance, autoencoders have been used to learn nonlinear latent representations from high-dimensional asset-pricing data \citep{gu2021}. In econometric forecasting, autoencoders have also been used as nonlinear dimension-reduction tools for large macroeconomic datasets \citep{hauzenberger2023}. These studies suggest that autoencoders provide a natural choice for extracting a nonlinear synthetic realized measure from a high-dimensional set of volatility proxies.

A complementary approach incorporates realized measure aggregation directly into Realized GARCH-type estimation. \citet{naimoli2022} consider adaptive RealEGARCH specifications in which the relative contribution of realized measures can depend on realized quarticity. This line of work shows that realized measure aggregation can be integrated with the volatility model rather than performed as a preliminary step. However, its focus on a small number of realized measures motivates parsimonious one-step methods that can incorporate a larger and more heterogeneous information set while retaining a single measurement equation.
Against this background, this paper develops three complementary approaches to incorporating multiple realized measure information into Realized GARCH. First, following the two-step structure of PC-RealGARCH and IC-RealGARCH, an autoencoder transforms a set of candidate realized measures into a single synthetic realized measure, which is subsequently incorporated into the Realized GARCH framework, yielding Autoencoder Enhanced Realized GARCH (AE-RealGARCH). Unlike PCA and ICA, the autoencoder can extract nonlinear common variation that is informative about volatility dynamics. AE-RealGARCH therefore incorporates richer realized measure information without introducing additional measurement equations, while retaining the established and interpretable structure of Realized GARCH.

Second, this paper proposes two one-step models that estimate realized measure aggregation jointly with the Realized GARCH parameters by maximum likelihood. Linear Weighted Realized GARCH (LW-RealGARCH) constructs a convex weighted average and estimates the contribution of each realized measure according to its fit within the volatility model. This differs from equal weighting and preliminary dimension reduction, where the combination is determined independently of the Realized GARCH likelihood. Functional Weighted Realized GARCH (FW-RealGARCH) instead parameterizes the aggregation weights using a mixture of two normalized Beta weighting functions \citep{ghysels2007}. The mixture involves four shape parameters and one mixing parameter, regardless of the number of candidate measures, allowing for flexible weight profiles while reducing reliance on the arbitrary ordering of realized measures. Thus, LW-RealGARCH prioritizes flexibility in measure-specific weighting, whereas FW-RealGARCH provides a parsimonious, function-based alternative as the realized measure dimension increases.

The contribution of this paper is threefold. Firstly, we extend synthetic realized measure construction from linear dimension reduction to nonlinear representation learning through an autoencoder. Secondly, we develop flexible linear and functional weighting schemes that jointly estimate realized measure aggregation and the Realized GARCH parameters via maximum likelihood. Thirdly, the paper provides a detailed and reproducible autoencoder design that compresses 12 realized measures, observed over approximately 22 years across 29 markets, into an informative synthetic realized measure. By extending autoencoder workflows from image-based research to financial time-series data, the accompanying code provides a template that can be readily adapted to higher-dimensional panels of realized measures. The code used to produce the empirical results is available at \url{https://github.com/...}\footnote{Github link anonymized for double anonymized peer review.}.

The remainder of this paper is organized as follows. Section \ref{sec:background_model} reviews the relevant background models. Section \ref{sec:proposed_models} introduces the proposed AE-RealGARCH, LW-RealGARCH, and FW-RealGARCH models. The following section, Section \ref{sec:model_estimation}, describes autoencoder training and likelihood estimation. Section \ref{sec:data_empirical_study} presents the data, forecasting design, and empirical results. Section \ref{sec:conclusion} concludes the paper.

\section{Background Model}\label{sec:background_model}
\subsection{Realized GARCH}\label{sec:realgarch}
The log version of Realized GARCH model of \citet{hansen2012} jointly models returns, latent conditional variance, and an observed realized measure. It consists of a return equation, a conditional variance equation, and a measurement equation:
\begin{equation}
\begin{aligned}
& r_t=\sigma_t z_t, \\
& \log(\sigma_t^2)=\omega+\beta \log(\sigma_{t-1}^2)+\gamma \log(x_{t-1}), \\
& \log(x_t)=\xi+\varphi \log(\sigma_t^2)+\tau_1 z_t+\tau_2\left(z_t^2-1\right)+\sigma_{\varepsilon} \varepsilon_t,
\label{eq:realgarch}
\end{aligned}
\end{equation}
where $r_t=[\log(C_t)-\log(C_{t-1})]\times 100$ is the percentage log return for day $t$, $z_t \overset{\mathrm{i.i.d.}}{\sim} \mathcal{N}(0,1)$ and $\varepsilon_t \overset{\mathrm{i.i.d.}}{\sim} \mathcal{N}(0,1)$ are mutually independent innovations, $\sigma_t^2$ is the conditional variance, and $x_t$ is a realized measure. The term $\tau_1 z_t+\tau_2(z_t^2-1)$ captures leverage effects in the measurement equation. In this baseline specification, the realized measure $x_t$ is one-dimensional. This feature is useful for parsimony, but it also motivates extensions that summarize or combine multiple realized measures before entering the Realized GARCH structure.

\citet{naimoli2022} construct synthetic realized measures using the first principal component from PCA and the first independent component from ICA, denoted by $x_{PC,t}$ and $x_{IC,t}$, respectively. They then apply min--max scaling to align these synthetic measures with the level of the original realized measures and replace $x_t$ in Equation~(\ref{eq:realgarch}) with $x_{PC,t}$ and $x_{IC,t}$, respectively, yielding the PC-RealGARCH and IC-RealGARCH models.

A simpler benchmark is AVG-RealGARCH, which uses the equally weighted average of the realized measures:
\begin{equation}
\bar{x}_t=\frac{1}{D_x}\sum_{d=1}^{D_x} x_{t,d}, \quad t=1,\ldots,T.
\label{eq:avg measure}
\end{equation}

Here, $x_{t,d}$ denotes the $d$-th realized measure at time $t$. AVG-RealGARCH is obtained by replacing $x_t$ in Equation~(\ref{eq:realgarch}) with $\bar{x}_t$. Thus, PC-RealGARCH, IC-RealGARCH, and AVG-RealGARCH preserve the Realized GARCH structure and differ only in how the realized measure input is constructed before model estimation. These models serve as benchmarks in this paper.

\subsection{Autoencoder}
An autoencoder is an unsupervised neural network that learns a low-dimensional representation by reconstructing its inputs \citep{hinton2006}. A basic single-hidden-layer autoencoder consists of an input layer, a lower-dimensional hidden layer, and an output layer. The encoder maps the high-dimensional input to a compact latent representation, while the decoder maps this representation back to the input space. Because the input also serves as the training target, no labeled data are required. The network parameters are estimated by minimizing the reconstruction error between the input and reconstructed output.

For an observation $\mathbf{x}_t \in \mathbb{R}^{D_x}$ at time $t$, the encoder and decoder mappings are defined as follows:
\begin{equation}
\begin{aligned}
& \mathbf{h}_t = f(\mathbf{W}^{(1)} \mathbf{x}_t + \mathbf{b}^{(1)}), \\
& \hat{\mathbf{x}}_t = g(\mathbf{W}^{(2)} \mathbf{h}_t + \mathbf{b}^{(2)}),
\end{aligned}
\label{eq:single_layer_ae}
\end{equation}
where $\mathbf{h}_t \in \mathbb{R}^{D^{(1)}}$ is the hidden-layer representation, with $D^{(1)}<D_x$. The weight matrices $\mathbf{W}^{(1)} \in \mathbb{R}^{D^{(1)}\times D_x}$ and $\mathbf{W}^{(2)} \in \mathbb{R}^{D_x\times D^{(1)}}$ map the input to the hidden layer and the hidden layer to the output, respectively. The vectors $\mathbf{b}^{(1)} \in \mathbb{R}^{D^{(1)}}$ and $\mathbf{b}^{(2)} \in \mathbb{R}^{D_x}$ are the corresponding bias terms.

The functions $f(\cdot)$ and $g(\cdot)$ are activation functions applied elementwise; common choices include the sigmoid and ReLU functions. The vector $\hat{\mathbf{x}}_t\in\mathbb{R}^{D_x}$ denotes the reconstructed input. Training selects the weights and biases to minimize the reconstruction loss over the full sample. For continuous inputs such as realized measures, a common loss function is the mean squared error (MSE), defined as
\begin{equation}
L = \frac{1}{T\times D_x} \sum_{t=1}^T \sum_{d=1}^{D_x} (x_{t,d} - \hat{x}_{t,d})^2,
\label{eq:mse equation}
\end{equation}
where $T$ is the sample size, $x_{t,d}$ is the value of the $d$-th realized measure at time $t$, and $\hat{x}_{t,d}$ is its reconstructed value.

\section{Proposed Models}
\label{sec:proposed_models}
\subsection{Autoencoder Enhanced Realized GARCH}
\label{sec:aerealgarch}
Following the sequential two-step procedure in PC-RealGARCH and IC-RealGARCH, we first use an autoencoder to compress the candidate realized measures into a single synthetic measure that captures their common volatility information. The resulting synthetic measure is then used as the realized measure input to the Realized GARCH model for volatility forecasting. This procedure yields AE-RealGARCH, which incorporates information from multiple realized measures while retaining the single realized measure input structure of Realized GARCH.

\begin{figure}[ht!]
    \centering
    \includegraphics[width=\linewidth]{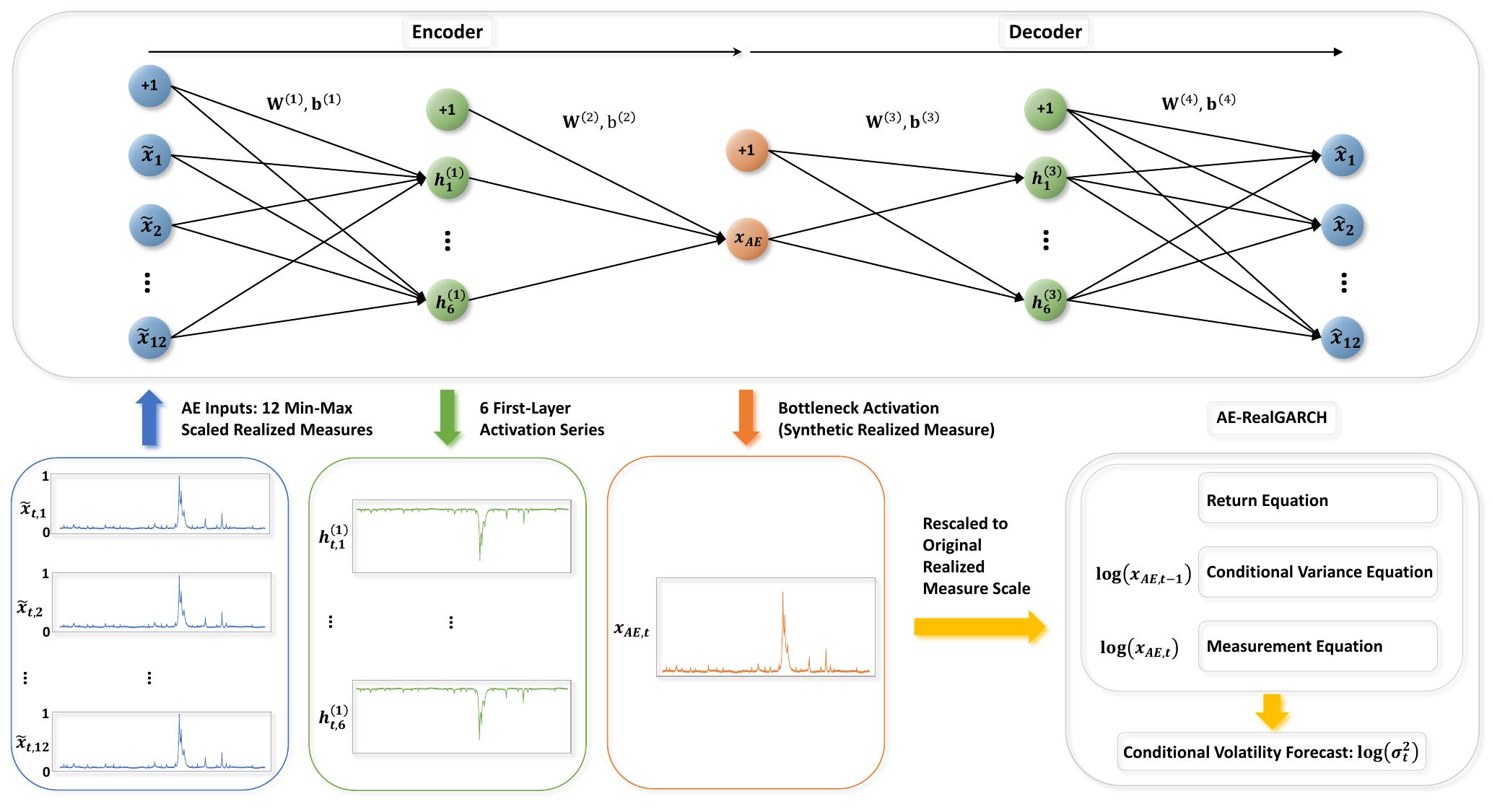}
    \caption{AE-RealGARCH architecture demonstration. The upper panel illustrates the 12-6-1-6-12 autoencoder used to construct the synthetic realized measure. The blue curves represent the min--max-scaled realized measure inputs $\tilde{x}_{t,d}$, $d=1,\ldots,12$; the green curves represent the activations of the first hidden layer, $h^{(1)}_{t,j}$, $j=1,\ldots,6$; the orange curve represents the bottleneck output $x_{AE,t}$. The lower-right panel illustrates how the rescaled bottleneck output enters the Realized GARCH framework.}
    \label{fig:ae_realgarch_structure}
\end{figure}

Figure~\ref{fig:ae_realgarch_structure} presents the architecture of the proposed AE-RealGARCH framework. The autoencoder contains five layers, and its weights and biases are trained by gradient descent using backpropagation. Each non-input layer applies the logistic sigmoid activation function, $\operatorname{sig}(u)=1/[1+\exp(-u)]$, elementwise. The input layer contains $D_x=12$ candidate realized measures. Before training, each raw realized measure $x_{t,d}$ is min--max scaled to obtain $\tilde{x}_{t,d}\in[0,1]$, and the resulting input vector is denoted by $\tilde{\mathbf{x}}_t\in\mathbb{R}^{D_x}$.

We consider two autoencoder architectures: 12-6-1-6-12 and 12-12-1-12-12. Figure~\ref{fig:ae_realgarch_structure} illustrates the 12-6-1-6-12 architecture. In this configuration, the 12-dimensional input is mapped to a six-neuron hidden layer and then compressed to a single bottleneck neuron, yielding $x_{AE,t}$. The decoder subsequently reconstructs the original input measures from this bottleneck representation. Training minimizes the regularized reconstruction error described in Section~\ref{sec:ae_training}. In the empirical analysis, the activations of the first hidden layer move inversely to the input realized measures, as can be seen from the patterns of the blue and green curves; Section~\ref{sec:insample_ae_encoded_series} discusses this important observation and its relation to regularization.

The forward pass of the autoencoder is given by:

\begin{equation}
\begin{aligned}
& \mathbf{h}^{(1)}_t = \operatorname{sig}(\mathbf{W}^{(1)}\tilde{\mathbf{x}}_t+\mathbf{b}^{(1)}), \\
& x_{AE,t} = \operatorname{sig}(\mathbf{W}^{(2)}\mathbf{h}^{(1)}_t+\mathbf{b}^{(2)}), \\
& \mathbf{h}^{(3)}_t = \operatorname{sig}(\mathbf{W}^{(3)}x_{AE,t}+\mathbf{b}^{(3)}), \\
& \hat{\mathbf{x}}_t = \operatorname{sig}(\mathbf{W}^{(4)}\mathbf{h}^{(3)}_t+\mathbf{b}^{(4)}),
\end{aligned}
\label{eq:multilayer_ae}
\end{equation}
where $\mathbf{h}^{(1)}_t$ and $\mathbf{h}^{(3)}_t$ denote the hidden-layer activations of the encoder and decoder, respectively, and $x_{AE,t}$ is the one-dimensional bottleneck representation. The sigmoid activation confines $x_{AE,t}$ to $(0,1)$, while regularization further compresses its variation. We therefore rescale $x_{AE,t}$ to match the empirical scale of the original realized measures before using it in the Realized GARCH model.

The AE-RealGARCH model is developed as follows:

\begin{equation}
\begin{aligned}
& r_t=\sigma_t z_t, \\
& \log(\sigma_t^2)=\omega+\beta \log(\sigma_{t-1}^2)+\gamma \log(x_{AE,t-1}), \\
& \log(x_{AE,t})=\xi+\varphi \log(\sigma_t^2)+\tau_1 z_t+\tau_2\left(z_t^2-1\right)+\sigma_{\varepsilon} \varepsilon_t.
\label{eq:ae-realgarch}
\end{aligned}
\end{equation}

\subsection{Linear Weighted Realized GARCH}\label{sec:lw_rgarch}
Unlike AE-RealGARCH, which constructs a synthetic realized measure before estimating the Realized GARCH parameters, LW-RealGARCH employs an one-step procedure. It forms a linearly weighted combination of the candidate realized measures and jointly estimates the aggregation weights and Realized GARCH parameters by maximizing a single likelihood function. The estimated weights, therefore, determine the combination of realized measures that best fits the model under the joint likelihood.

The LW-RealGARCH model is defined as follows:

\begin{equation}
\begin{aligned}
& r_t=\sigma_t z_t, \\
& \log(\sigma_t^2)=\omega+\beta \log(\sigma_{t-1}^2)+\gamma \log(x_{LW,t-1}), \\
& x_{LW,t}=\sum_{d=1}^{D_x} w_d x_{t,d}, \\
& \log(x_{LW,t})=\xi+\varphi \log(\sigma_t^2)+\tau_1 z_t+\tau_2\left(z_t^2-1\right)+\sigma_{\varepsilon} \varepsilon_t,
\label{eq:lw-realgarch}
\end{aligned}
\end{equation}
where $x_{t,d}$ denotes the $d$-th candidate realized measure and $w_d$ denotes its corresponding weight. The constraints $\sum_{d=1}^{D_x}w_d=1$ and $w_d\geq0$, for $d=1,\ldots,D_x$, ensure that $x_{LW,t}$ is a convex combination. The resulting aggregate enters the conditional variance and measurement equations as a single realized measure input, thereby preserving the single-measure structure of Realized GARCH.

Variation in the estimated weights across estimation windows shows how the relative importance of the candidate realized measures changes over time. Sections~\ref{sec:insmp_w_lwrealgarch} and~\ref{sec:dynamic_weight_behavior} examine the in-sample and out-of-sample behavior of these weights, respectively.

\subsection{Functional Weighted Realized GARCH}\label{sec:nlw_rgarch}
We extend the linear weighting approach in LW-RealGARCH by using a function to generate the weight assigned to each realized measure. We refer to this specification as Functional Weighted Realized GARCH (FW-RealGARCH). The weight profile is determined by a small set of function parameters, reducing the number of aggregation parameters compared with estimating measure-specific weights in LW-RealGARCH.

We parameterize the weight profile using the Beta polynomial developed in the mixed data sampling (MIDAS) literature \citep{ghysels2007}. Following the application of MIDAS-style weighting to volatility forecasting by \citet{storti2022}, we use a mixture of two normalized Beta weighting functions to aggregate the realized measures. Each Beta function depends on two shape parameters, $a$ and $b$. For component $k=1,2$, the normalized weights are defined as
\begin{equation}
w_d^{(k)}=\frac{\left(\frac{d}{D_x+1}\right)^{a_k-1}\left(1-\frac{d}{D_x+1}\right)^{b_k-1}}{\sum_{j=1}^{D_x}\left(\frac{j}{D_x+1}\right)^{a_k-1}\left(1-\frac{j}{D_x+1}\right)^{b_k-1}},\quad k=1,2,
\label{eq:fw-beta-components}
\end{equation}
where $a_k>0$ and $b_k>0$ are the shape parameters of component $k$. A single Beta weighting function is commonly applied to lagged variables, whose ordering is determined by time. However, the realized measures considered here are observed on the same day and have no natural temporal ordering. To address this limitation, we employ a mixture of two Beta weighting functions, allowing for a more flexible weighting profile and reducing reliance on the arbitrary ordering of the realized measures.

The FW-RealGARCH model is defined as follows:

\begin{equation}
\begin{aligned}
& r_t = \sigma_t z_t, \\
& \log(\sigma_t^2) = \omega + \beta \log(\sigma_{t-1}^2) + \gamma \log(x_{FW,t-1}), \\
& x_{FW,t} = \sum_{d=1}^{D_x} w_d x_{t,d}, \\
& w_d = \eta w_d^{(1)}+(1-\eta)w_d^{(2)}, \\
& \log(x_{FW,t}) = \xi + \varphi \log(\sigma_t^2) + \tau_1 z_t + \tau_2(z_t^2-1) + \sigma_{\varepsilon} \varepsilon_t,
\end{aligned}
\label{eq:nln-realgarch}
\end{equation}
where $\eta\in[0,1]$ is the mixing parameter. The resulting weights satisfy $w_d\geq0$ and $\sum_{d=1}^{D_x}w_d=1$, so $x_{FW,t}$ remains a convex combination of the candidate realized measures.

Appendix Figure~\ref{fig:appendix_beta_weights} illustrates the shapes of individual Beta profiles. A component is flat when $a_k=b_k=1$, decreases when $a_k=1$ and $b_k>1$, and increases when $a_k>1$ and $b_k=1$. When the parameters of both Beta components are jointly estimated with the Realized GARCH parameters, their mixture can capture a wide range of flexible, nonlinear weight profiles. Section~\ref{sec:insmp_w_nlwrealgarch} discusses the estimated in-sample weight profiles, and Section~\ref{sec:dynamic_weight_behavior} examines how they evolve across rolling estimation windows. 

Future work could also explore the relative-score combination approach of \citet{taylor2020a} as an alternative weighting scheme, allowing aggregation weights to respond more directly to volatility or risk forecasting performance.

\section {Model Estimation}
\label{sec:model_estimation}
The estimation procedure depends on how the candidate realized measures are aggregated. LW-RealGARCH and FW-RealGARCH are estimated in a single step by maximum likelihood estimation (MLE). For LW-RealGARCH, the linear weights $w_d$ are estimated jointly with the standard Realized GARCH parameters $(\omega,\beta,\gamma,\xi,\varphi,\tau_1,\tau_2,\sigma_{\varepsilon})$. For FW-RealGARCH, the mixing parameter $\eta$ and the four Beta shape parameters $(a_1,b_1,a_2,b_2)$ are estimated jointly with the same Realized GARCH parameters.

AE-RealGARCH follows a sequential two-step procedure. First, the autoencoder is trained to construct the nonlinear synthetic realized measure $x_{AE,t}$. Second, this measure is treated as the observed realized measure input, and the standard Realized GARCH parameters are then estimated by MLE.

\subsection{Autoencoder Training}
\label{sec:ae_training}

We train the multi-layer autoencoder by minimizing an objective that combines the MSE reconstruction loss with weight and sparsity regularization. Following the sparse autoencoder formulation described by \citet{themathworksinc.2026}, the loss function is
\begin{equation}
\label{eq:loss_with_regularisation}
L=
\underbrace{\frac{1}{T\times D_x}\sum_{t=1}^T\sum_{d=1}^{D_x}
\left(\tilde{x}_{t,d}-\hat{x}_{t,d}\right)^2}_{\text{MSE}}
+\lambda_1\underbrace{\mathbf{\Omega}_{\text{weights}}}_{\text{weight regularization}}
+\lambda_2\underbrace{\mathbf{\Omega}_{\text{sparsity}}}_{\text{sparsity regularization}},
\end{equation}
where $\lambda_1$ and $\lambda_2$ control the strengths of the weight and sparsity regularization terms, respectively. The two components are defined below.

\subsubsection{Weight Regularization}
\label{sec:weight_regularization}
The candidate realized measures proxy the same latent volatility and are therefore highly correlated despite their different construction procedures. An $L_1$ penalty encourages sparse solutions and may select only a subset of these correlated inputs. We instead use an $L_2$ penalty, which shrinks the encoder weights toward zero without imposing exact zeros. This controls the magnitude of the network weights while retaining information across correlated realized measures. The weight regularization term is

\begin{equation}
\mathbf{\Omega}_{\text{weights}} = \frac{1}{2}( \|\mathbf{W}^{(1)}\|_F^2 + \|\mathbf{W}^{(2)}\|_F^2),
\end{equation}
where $\|\cdot\|_F$ denotes the Frobenius norm, and $\mathbf{W}^{(1)}$ and $\mathbf{W}^{(2)}$ are the weight matrices associated with the first and second encoder layers, respectively.

\subsubsection{Sparsity Regularization}
\label{sec:sparsity_regularization}
Weight regularization limits the magnitude of the encoder weights but does not directly constrain the activation patterns of the hidden layers. During training, we found that an $L_2$ penalty alone could produce latent series with level shifts and short-lived fluctuations that were not well aligned with the common movements of the input realized measures. Motivated by this empirical behavior and the sparse-representation approach of \citet{olshausen1997a}, we add a sparsity penalty to encourage a more parsimonious latent representation.

The sparsity penalty uses the Kullback--Leibler (KL) divergence to compare the average activation of each encoder unit with a target level, $\rho$. It discourages units from remaining persistently active and can reduce the influence of short-lived, measure-specific variation on the encoded representation. In this application, the penalty is intended to help the bottleneck retain the common and persistent volatility patterns shared across the candidate realized measures. The complete sparsity penalty applied across the encoder layers is
\begin{equation}
\mathbf{\Omega}_{\text{sparsity}}
=
\sum_{j=1}^{D^{(1)}}\operatorname{KL}\left(\rho\middle\|\hat{\rho}_j^{(1)}\right)
+
\sum_{j=1}^{D^{(2)}}\operatorname{KL}\left(\rho\middle\|\hat{\rho}_j^{(2)}\right),
\end{equation}
where $D^{(1)}$ and $D^{(2)}$ denote the numbers of units in the first and second encoder layers, respectively. The empirical average activation of unit $j$ in layer $l$ over the training sample is
\begin{equation}
\hat{\rho}_j^{(l)}
=
\frac{1}{T}\sum_{t=1}^{T}h_{t,j}^{(l)},
\end{equation}
where $h_{t,j}^{(l)}$ denotes the sigmoid activation of unit $j$ in encoder layer $l$ for the $t$-th observation. The deviation of $\hat{\rho}_j^{(l)}$ from $\rho$ is measured using the Kullback--Leibler divergence:
\begin{equation}
\operatorname{KL}\left(\rho\middle\|\hat{\rho}_j^{(l)}\right)
=
\rho\log\left(\frac{\rho}{\hat{\rho}_j^{(l)}}\right)
+
(1-\rho)\log\left(\frac{1-\rho}{1-\hat{\rho}_j^{(l)}}\right).
\end{equation}

Sparse autoencoders commonly use a fixed target activation. However, the distributions of the min--max-scaled realized measures vary across markets and training windows. Based on robustness checks across alternative values of $\rho$, we define $\rho$ as the pooled mean of the min--max-scaled realized measures within each training sample:
\begin{equation}
\rho
=
\frac{1}{T\times D_x}
\sum_{t=1}^{T}\sum_{d=1}^{D_x}\tilde{x}_{t,d}.
\label{eq:target_rho}
\end{equation}

Because $\tilde{x}_{t,d}\in[0,1]$, the resulting value of $\rho$ lies on the same scale as the sigmoid activations. This adaptive definition guides the average activation $\hat{\rho}_j^{(l)}$ toward a meaningful, sample-specific target and can help stabilize the latent series across markets and training windows. The detailed training configuration and implementation settings are reported in Appendix~\ref{app:ae_training_config}, to ensure the reproducibility of the autoencoder training procedure.

\subsection{Maximum Likelihood Estimation of the Realized GARCH Framework}
We estimate the proposed models using maximum likelihood estimation. Under the Gaussian innovation specification in Section~\ref{sec:realgarch}, the measurement-equation residual is $\mu_t=\sigma_\varepsilon\varepsilon_t$, where $\mu_t \overset{\mathrm{i.i.d.}}{\sim}\mathcal{N}(0,\sigma_\varepsilon^2)$. Omitting constant terms, the joint log-likelihood for Models~\ref{eq:ae-realgarch}, \ref{eq:lw-realgarch}, and \ref{eq:nln-realgarch} is

\begin{equation}
\label{eq:ll_realgarch}
l(\mathbf{r,x}; \boldsymbol{\theta})=\underbrace{-\sum_{t=1}^{T} \left[\log(\sigma_{t}^2)+\frac{r^2_{t}}{\sigma_{t}^2}\right]}_{l(\mathbf{r};\boldsymbol{\theta})}\underbrace{-\sum_{t=1}^{T}\left[\log(\sigma_{\varepsilon}^2)+\frac{\mu^2_{t}}{\sigma_{\varepsilon}^2}\right]}_{l(\mathbf{x}\mid\mathbf{r};\boldsymbol{\theta})},
\end{equation}
where
$
\mu_t
=
\log(x_t)-\xi-\varphi\log(\sigma_t^2)-\tau_1z_t-\tau_2\left(z_t^2-1\right),
$
and $\boldsymbol{\theta}$ denotes the model-specific parameter vector. For LW-RealGARCH, $\boldsymbol{\theta}$ contains the 12 linear weights, subject to the convexity constraints, together with the standard Realized GARCH parameters:
$\boldsymbol{\theta}
=
[w_1,\ldots,w_{12},\omega,\beta,\gamma,\xi,\varphi,\tau_1,\tau_2,\sigma_\varepsilon]^{\intercal}.$
For FW-RealGARCH, $\boldsymbol{\theta}$ contains the five aggregation parameters and the eight standard Realized GARCH parameters:
$\boldsymbol{\theta}
=
[\eta,a_1,b_1,a_2,b_2,\allowbreak
\omega,\beta,\gamma,\xi,\varphi,\allowbreak
\tau_1,\tau_2,\sigma_\varepsilon]^{\intercal}.$
For AE-RealGARCH, the autoencoder is trained in the preceding step. The likelihood estimation procedure therefore only estimates the standard Realized GARCH parameters conditional on the synthetic measure:
$\boldsymbol{\theta}
=
[\omega,\beta,\gamma,\xi,\varphi,\tau_1,\tau_2,\sigma_\varepsilon]^{\intercal}.$
Section~\ref{sec:insample_params_etm} reports and discusses the resulting in-sample parameter estimates. It is worth noting that, although the Gaussian distribution is adopted in this study, heavy-tailed and skewed distributions, such as the Student's t and skewed-t distributions, could be incorporated to further improve model performance.

\section{Data and Empirical Study}\label{sec:data_empirical_study}
\subsection{Data Description}
The empirical analysis uses data from the Oxford-Man Institute's Realised Library \citep{gerdheber2009}. The dataset provides 12 candidate realized measures for each of 29 global equity indices, including the S\&P 500, FTSE, and Nikkei 225 (N225). The sample spans January 2000 to June 2022, and daily returns are computed from closing prices. Each market series excludes market-specific non-trading days in accordance with its local trading calendar. Most indices contain approximately 5,500 daily observations; FTSE MIB (FTMIB) is the main exception, with 3,314 observations. Table~\ref{tab:rt_stats} reports descriptive statistics for the return series in each market.

The sample is divided approximately 80:20 into in-sample and out-of-sample periods. The in-sample period, from January 2000 to December 2017, includes the 2008 Global Financial Crisis. The out-of-sample forecasting period, from January 2018 to June 2022, covers the market disruption associated with the COVID-19 pandemic. We also conduct a separate COVID-period analysis, from 30 January 2020 to 30 June 2020, to assess the performance of synthetic realized measures during a high-volatility episode. This window begins on the date when the World Health Organization declared the outbreak a public health emergency of international concern and includes 11 March 2020, when the outbreak was characterized as a pandemic \citep{who2026}.

\input{Table/rt_stats}

\subsection{Forecasting Set-up}
\label{sec:forecasting_setup}
To evaluate out-of-sample forecasting performance, we use a fixed-length rolling window to generate one-step-ahead volatility forecasts. Using the initial estimation sample, $t=1,\ldots,T_{\text{in}}$, we estimate the parameter vector $\hat{\boldsymbol{\theta}}$ and produce the conditional variance forecast $\hat{\sigma}_{T_{\text{in}}+1\mid T_{\text{in}}}^2$. We then advance both endpoints of the estimation window by one trading day, re-estimate the model, and generate the next forecast. The window length remains $T_{\text{in}}$, and this procedure is repeated over the full out-of-sample period.

We compare the proposed models with four established benchmark frameworks: PC-RealGARCH, IC-RealGARCH, AVG-RealGARCH, and the standard Realized GARCH model, as reviewed in Section~\ref{sec:background_model}. To assess performance during periods of severe market stress, we evaluate forecasts over both the full out-of-sample period and the COVID period. We use three evaluation criteria: predictive log-likelihood for volatility forecasts, a quantile score for VaR, and a joint loss for VaR and ES.

First, we assess volatility forecasts using the partial predictive log-likelihood (PLL). This criterion evaluates how well the conditional variance forecast explains the realized out-of-sample return. Omitting the constant term, the aggregate PLL is given by

\begin{equation}
\label{eq:pnll_garch_part}
\text{PLL}:= -\sum_{t=T_{\text{in}}+1}^{T_{\text{in}}+T_{\text{out}}}\left[\log(\hat{\sigma}_{t}^2)+\frac{r^2_{t}}{\hat{\sigma}_{t}^2}\right].
\end{equation}

In the empirical results reported in Section~\ref{sec:outsample_fc_pfm}, we use its negative, NPLL, so that lower values indicate better volatility forecasts. 

Second, to assess tail-risk forecasts, we compute one-step-ahead forecasts of daily VaR and ES, denoted by $\widehat{\mathrm{VaR}}_t$ and $\widehat{\mathrm{ES}}_t$, at probability level $\alpha$. The $\alpha$-level VaR is the conditional $\alpha$-quantile of returns, whereas ES is the conditional expected return given a realization at or below the VaR threshold. In line with the Basel market risk framework \citep{baselcommitteeonbankingsupervision2019}, we set $\alpha=2.5\%$. Because quantiles are elicitable risk measures \citep{gneiting2011}, we evaluate VaR forecasts using the strictly consistent quantile score. We report its sample average over the out-of-sample period:

\begin{equation}
\label{eq:qloss}
\mathrm{QS}:= \frac{1}{T_{\text{out}}}
\sum_{t=T_{\text{in}}+1}^{T_{\text{in}}+T_{\text{out}}}
\left( \alpha - \mathit{I}\{ r_t \leq \widehat{\mathrm{VaR}}_t \} \right)
\left( r_t - \widehat{\mathrm{VaR}}_t \right),
\end{equation}
where $\mathit{I}\{\cdot\}$ equals one when $r_t$ is at or below the forecasted VaR and zero otherwise. Lower quantile score values indicate more accurate VaR forecasts.

Third, ES is not individually elicitable \citep{gneiting2011}. However, \citet{fissler2016} show that VaR and ES are jointly elicitable. Building on this result, \citet{taylor2019} show that the negative log-likelihood of the Asymmetric Laplace (AL) distribution provides a strictly consistent joint scoring function for VaR and ES. We therefore evaluate joint VaR--ES forecast accuracy using the AL joint loss:

\begin{equation}
\label{eq:jointloss}
\mathrm{AL}:= \frac{1}{T_{\text{out}}} \sum_{t=T_{\text{in}}+1}^{T_{\text{in}}+T_{\text{out}}} \left( -\log\left(\frac{\alpha - 1}{\widehat{\mathrm{ES}}_t}\right) - \frac{(r_t - \widehat{\mathrm{VaR}}_t)(\alpha - \mathit{I}\{{r_t \leq \widehat{\mathrm{VaR}}_t}\})}{\alpha \widehat{\mathrm{ES}}_t} \right),
\end{equation}
where lower AL joint loss values indicate more accurate joint VaR--ES forecasts.

\subsection{In-Sample Study}
\subsubsection{Autoencoder Hidden-Layer and Bottleneck Activations}
\label{sec:insample_ae_encoded_series}
Following the AE-RealGARCH framework introduced in Section \ref{sec:aerealgarch}, this section examines the in-sample representations learned by the 12-6-1-6-12 autoencoder for the S\&P 500. We first analyze the first hidden-layer activation series and then examine the bottleneck activation series. This helps illustrate how the information contained in the 12 realized measures is transformed into a single synthetic realized measure for subsequent use in the Realized GARCH model.

\begin{figure}[ht!]
    \centering
    \includegraphics[width=0.8\linewidth]{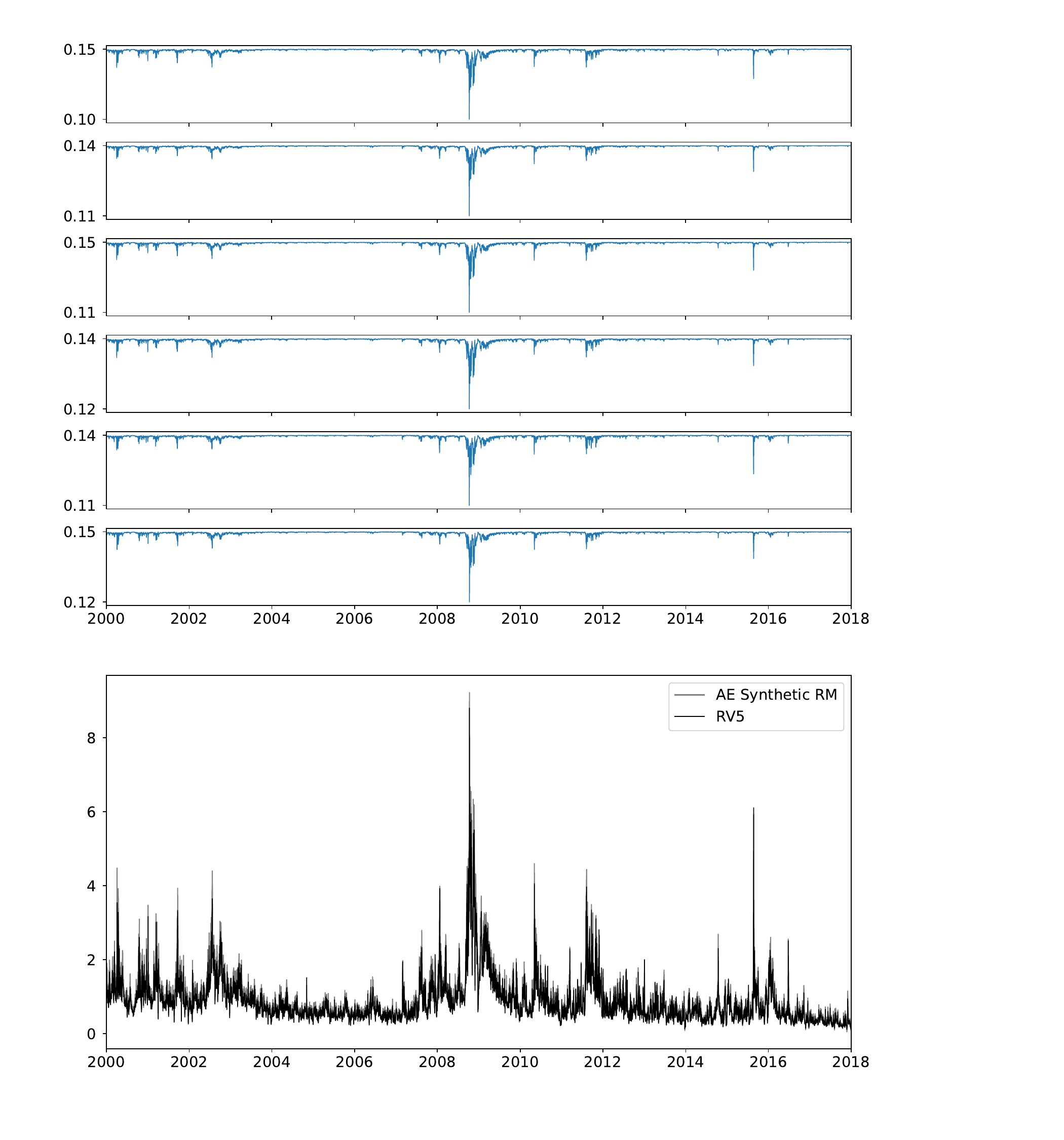}
    \caption{The upper six plots display the first hidden-layer activation series. The bottom plot compares the rescaled synthetic realized measure (AE Synthetic RM) with RV5, with both series expressed on the volatility scale.}
    \label{fig:ae_acts_and_synrm}
\end{figure}

The upper six plots of Figure \ref{fig:ae_acts_and_synrm} reports the six first hidden-layer activation series. They correspond to the conceptual first-layer activations (in green) in the AE-RealGARCH architecture shown in Figure \ref{fig:ae_realgarch_structure}. The activations move inversely with the realized measures. Despite this inverse direction, the pronounced troughs in the activation series align with the major peaks in 5-minute realized variance (RV5) shown in the bottom panel, particularly over 2000--2003, 2008--2009, 2010--2012, and 2016, indicating that the encoded series preserve the timing of major volatility episodes. 

Under the regularized training procedure described in Section~\ref{sec:weight_regularization}, the estimated encoder weights are predominantly negative. Because the sigmoid activation function is monotonic, larger values of the min--max scaled realized measures tend to reduce the hidden-layer activations, generating the observed inverse relationship. Appendix Figure~\ref{fig:appendix_encoder_wts_hm} displays the corresponding encoder weights and confirms this negative weight structure.

This inverse relationship is not confined to a single neuron or sample period. In our analysis, it appears across all hidden neurons, markets, and rolling-window estimation steps. This regularity supports the use of an intermediate hidden layer in the 12-6-1 and 12-12-1 encoders, rather than directly compressing the 12-dimensional input into a single neuron. The first hidden layer captures a common inverse representation of the realized measures, while the bottleneck layer applies a further nonlinear transformation that reverses this direction and produces a series more closely aligned with the original realized measures.

The sparsity regularization described in Section~\ref{sec:sparsity_regularization} also affects the scale of the activation series. The KL-divergence penalty encourages the average activation of each hidden neuron to remain close to the adaptive target value $\rho$ defined in Equation~(\ref{eq:target_rho}). Therefore, this constraint places the activations within a relatively narrow range, approximately between 0.10 and 0.15.

The bottom plot in Figure~\ref{fig:ae_acts_and_synrm} compares the rescaled synthetic realized measure with RV5. The synthetic realized measure corresponds to the bottleneck activation series (in orange) in the AE-RealGARCH architecture shown in Figure \ref{fig:ae_realgarch_structure}. Because the sigmoid activation function and regularization compress the bottleneck activation into a limited range, we rescale it to the empirical scale of the original realized measures before using it in the Realized GARCH model.

After rescaling, the synthetic realized measure closely tracks RV5. This co-movement suggests that the bottleneck representation retains the common volatility information contained in the original realized measures. The rescaled bottleneck series can therefore serve as a meaningful synthetic proxy for latent volatility.

\subsubsection{Estimated Weights in LW-RealGARCH}\label{sec:insmp_w_lwrealgarch}
\begin{figure}[htbp]
    \centering
    \includegraphics[width=\linewidth]{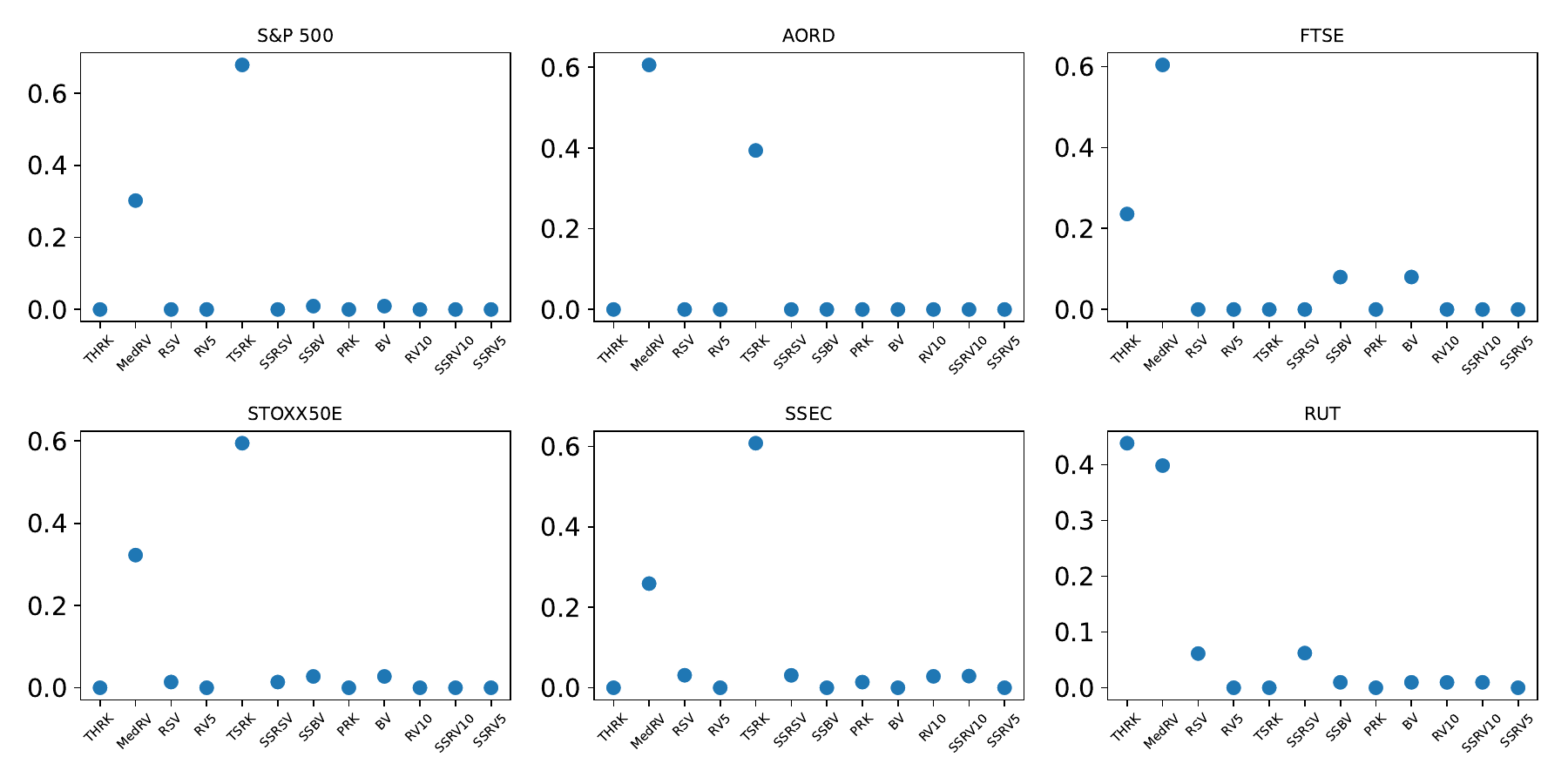}
    \caption{In-sample LW-RealGARCH weights across markets. The 12 candidate realized measures are THRK, MedRV, RSV, RV5, TSRK, SSRSV (subsampled realized semivariance), SSBV (subsampled bipower variation), PRK (realized kernel with a non-flat Parzen window), BV, RV10 (10-minute realized variance), SSRV10 (subsampled 10-minute realized variance), and SSRV5 (subsampled 5-minute realized variance).}
    \label{fig:lw_weights}
\end{figure}

As discussed in Section~\ref{sec:lw_rgarch}, LW-RealGARCH jointly estimates a set of measure-specific weights with the standard Realized GARCH parameters. Figure~\ref{fig:lw_weights} reports the in-sample weights assigned to the 12 candidate realized measures for six major indices: the S\&P 500, FTSE, Australian All Ordinaries (AORD), EURO STOXX 50 (STOXX50E), Shanghai Composite (SSEC), and Russell 2000 (RUT). These indices span North America, Europe, and the Asia-Pacific region, illustrating how the contributions of the candidate measures vary across markets.

Out of the 12 measures, the estimated weights are concentrated on three measures: MedRV \citep{andersen2012}, the realized kernel with a Tukey-Hanning window (THRK) \citep{barndorff-nielsen2008}, and the two-scale realized kernel (TSRK) \citep{ikeda2015}. These preferred measures are highly robust to market microstructure noise and price jumps. Specifically, MedRV mitigates the impact of infrequent price jumps while reducing sensitivity to extreme return outliers. Furthermore, THRK and TSRK are explicitly designed to counteract market microstructure frictions, such as bid-ask bounce and asynchronous trading.

Accordingly, for the S\&P 500, AORD, STOXX50E, and SSEC, the estimated synthetic realized measure is a linear combination dominated by MedRV and TSRK. This combination places most of the weight on measures designed to mitigate jump contamination and market microstructure noise. For the FTSE and RUT, the corresponding linear combination is instead driven primarily by MedRV and THRK. The estimated weight profiles, therefore, show how LW-RealGARCH constructs a market-specific realized measure by linearly combining complementary volatility estimators, rather than relying on a single preselected measure. The out-of-sample evolution of these weights is examined in Section~\ref{sec:dynamic_weight_behavior}.

\subsubsection{Estimated Weights in FW-RealGARCH}\label{sec:insmp_w_nlwrealgarch}
\begin{figure}[htbp]
    \centering
    \includegraphics[width=\linewidth]{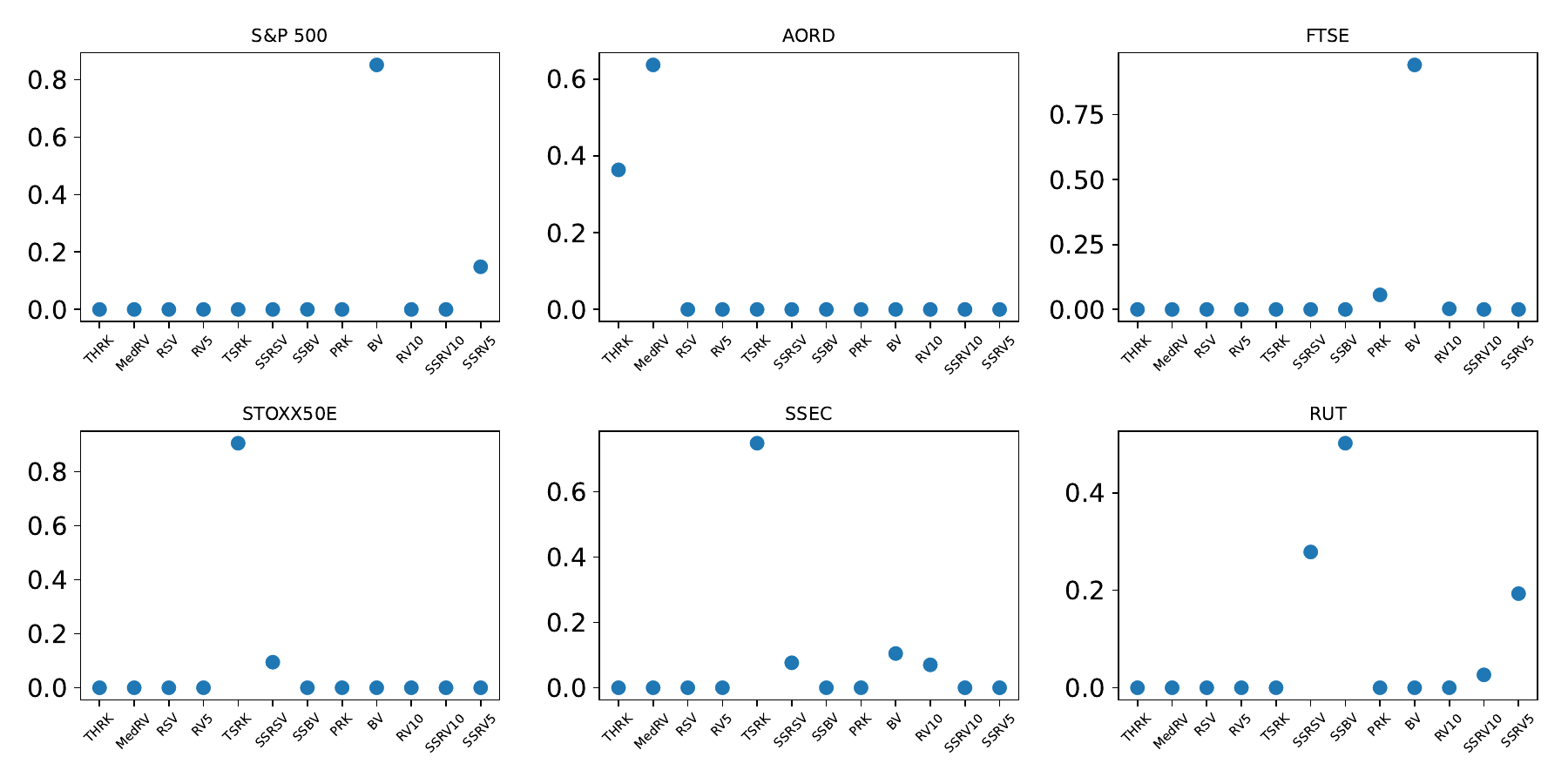}
    \caption{In-sample FW-RealGARCH weights across markets.}
    \label{fig:nlw_weights}
\end{figure}

As discussed in Section~\ref{sec:nlw_rgarch}, FW-RealGARCH generates measure weights using a mixture of two Beta weighting functions. Compared with a single Beta profile, this mixture allows more varied combinations of realized measures while retaining a parsimonious representation. Figure~\ref{fig:nlw_weights} presents the estimated in-sample weights and illustrates market-specific weighting patterns.

First, several preferred measures, including THRK, MedRV, and TSRK, are also prominent in the LW-RealGARCH results. AORD places most weight on a combination of THRK and MedRV, with the preference for MedRV consistent with LW-RealGARCH. The agreement is more evident in STOXX50E and SSEC, where TSRK receives the largest weight under both specifications.

Second, the S\&P 500 and FTSE exhibit profiles dominated by BV, while the S\&P 500 also assigns weight to subsampled realized variance at the 5-minute frequency (SSRV5). BV measures continuous price variation while remaining robust to infrequent jumps in the log-price process \citep{barndorff-nielsen2004}. These properties provide a possible explanation for the preference for BV in these two markets.

Third, the preference for subsampled measures is more evident in RUT, where the largest weights are assigned to SSBV and subsampled realized semivariance (SSRSV), followed by SSRV5, with a smaller contribution from SSRV10. Subsampling averages estimates across different sampling grids, reducing sensitivity to a particular choice of sampling times and making fuller use of the available high-frequency observations \citep{zhang2005}. The estimated weights suggest that these subsampled measures provide useful complementary information for constructing the synthetic realized measure in RUT.

FW-RealGARCH links individual weights through a mixture of Beta components, allowing more varied combinations of measures while showing some consistency with LW-RealGARCH. The estimated profiles show no clear order-dependent pattern of the kind typically associated with a single Beta weighting function. Notably, standard high-frequency estimators such as RV5 receive negligible weight under both modeling frameworks. This consistent down-weighting highlights the limitations of unadjusted RV5 when it is assessed against more microstructure-robust alternatives. Section \ref{sec:outsample_fc_pfm} further demonstrates the empirical advantage of using a synthesized volatility proxy rather than relying on a single traditional measure such as RV5.

\subsubsection{In-Sample Synthetic Realized Measures and Parameter Estimates}
\label{sec:insample_params_etm}
Table~\ref{tab:spx_log_rm_stats} reports descriptive statistics for the in-sample log-transformed realized measures. The synthetic realized measures have broadly similar dispersion to RV5, although their mean levels differ. The two autoencoder-based measures have the highest and similar means ($-0.418$ and $-0.456$) and standard deviations ($1.179$ and $1.176$). By contrast, the LW measure has the lowest mean ($-0.939$) and the smallest standard deviation ($1.105$). The PCA- and ICA-based measures have identical descriptive statistics in this sample.

The two autoencoder-based measures and the PCA- and ICA-based measures share the same minimum and maximum values because they are min--max rescaled to the empirical range of the realized measures. They also exhibit lower skewness than the weighting-, averaging-, and RV5-based measures. For example, the skewness values of the two autoencoder-based measures are $0.186$ and $0.193$, compared with $0.340$ for the averaged realized measure. Thus, the dimension-reduction methods produce less skewed synthetic-measure distributions in the in-sample period.

\input{Table/spx_log_rm_stats}
Table~\ref{tab:insample_params_spx} reports the in-sample parameter estimates for the proposed and benchmark Realized GARCH specifications. Overall, the parameter estimates are broadly comparable across models. The measurement-equation coefficient $\varphi$ is close to one in all specifications, while $\tau_1$, $\tau_2$, and $\sigma_\varepsilon$ remain similar in magnitude. This suggests that the leverage effect and measurement-error variance are relatively stable across different specifications. The two AE-RealGARCH specifications also yield nearly identical parameter estimates, indicating that the choice between the 12-6-1 and 12-12-1 architectures has little effect on the in-sample estimates.

Nevertheless, several differences in Table~\ref{tab:insample_params_spx} are worth noting. First, $\xi$ is substantially higher, or less negative, in the AE-RealGARCH models ($-0.215$ and $-0.252$) than in the other Realized GARCH-type models, such as AVG-RealGARCH ($-0.646$). Since $\xi$ represents the fixed bias component in the measurement equation, its value being closer to zero suggests that the log-transformed autoencoder-based measure is closer to serving as a proxy for the log-conditional variance. Correspondingly, the intercept $\omega$ in the conditional variance equation is smaller for the AE-RealGARCH models ($0.073$ and $0.087$) than for AVG-RealGARCH ($0.242$). This offset becomes clearer when the model is written in its autoregressive representation of order one, $\log(\sigma^2_t)=(\omega+\gamma \xi)+\pi\log(\sigma^2_{t-1})+\gamma w_{t-1}$, where $\pi=\beta+\gamma\varphi$ and $w_{t-1}=\tau_1 z_{t-1}+\tau_2(z_{t-1}^2-1)+\mu_{t-1}$. In this representation, $\omega+\gamma\xi$ is the effective intercept. The combined coefficients remain comparable across models, indicating that the less negative $\xi$ in AE-RealGARCH largely compensates for its lower $\omega$. Second, the Realized GARCH model using RV5 as the realized measure has a smaller $\gamma$ ($0.357$) than models based on synthetic realized measures, such as LW-RealGARCH ($0.476$). This pattern is consistent with the possibility that, synthetic realized measures provide a stronger lagged signal for conditional variance than RV5.

To assess in-sample fit, we also report the training negative log-likelihood (NLL) in Table~\ref{tab:insample_params_spx}, computed as the negative of the return-likelihood component $l(\mathbf{r};\boldsymbol{\theta})$ defined in Equation~(\ref{eq:ll_realgarch}). Lower NLL values indicate a better in-sample fit of the conditional return distribution. Among the two-step specifications, the AE-RealGARCH models achieve lower NLL values ($3671.0$ and $3672.0$) than PC-RealGARCH and IC-RealGARCH ($3687.6$ for both), as well as the Realized GARCH model based on RV5 ($3683.9$). These results indicate that the autoencoder-based synthetic measures provide a better in-sample fit than the PCA-, ICA-, and RV5-based measures.

AVG-RealGARCH serves as a competitive benchmark by using the averaged realized measure, with an NLL of $3665.9$. However, LW-RealGARCH and FW-RealGARCH achieve lower NLL values ($3663.3$ and $3644.5$, respectively), indicating that the estimated weighting schemes provide a closer in-sample fit than simple averaging. Among all specifications, FW-RealGARCH attains the lowest NLL ($3644.5$), indicating the best in-sample fit.

The `Time' row in Table \ref{tab:insample_params_spx} reports the average seconds required for realized measure processing, model fitting, and the first volatility forecast. For one-step models, such as LW-RealGARCH and FW-RealGARCH, the reported time covers the full joint estimation and forecasting procedure. For two-step models, such as AE-RealGARCH, the reported time is decomposed into realized measure processing time plus Realized GARCH fitting and forecasting time. Constructing the autoencoder-based measure requires only $0.143$ and $0.141$ seconds, respectively, while the subsequent Realized GARCH estimation requires approximately $0.1$ seconds for each architecture. The additional autoencoder processing time is modest in absolute terms and remains acceptable. LW-RealGARCH and FW-RealGARCH require approximately $0.4$ and $0.1$ seconds, respectively, for joint estimation of the weighting and Realized GARCH parameters and the first volatility forecast.

\input{Table/insample_params_v2}

\subsection{Out-of-Sample Study}
\subsubsection{Dynamic Weight Behavior in LW-RealGARCH and FW-RealGARCH}
\label{sec:dynamic_weight_behavior}
\begin{figure}[htbp]
    \centering
    \includegraphics[width=\linewidth]{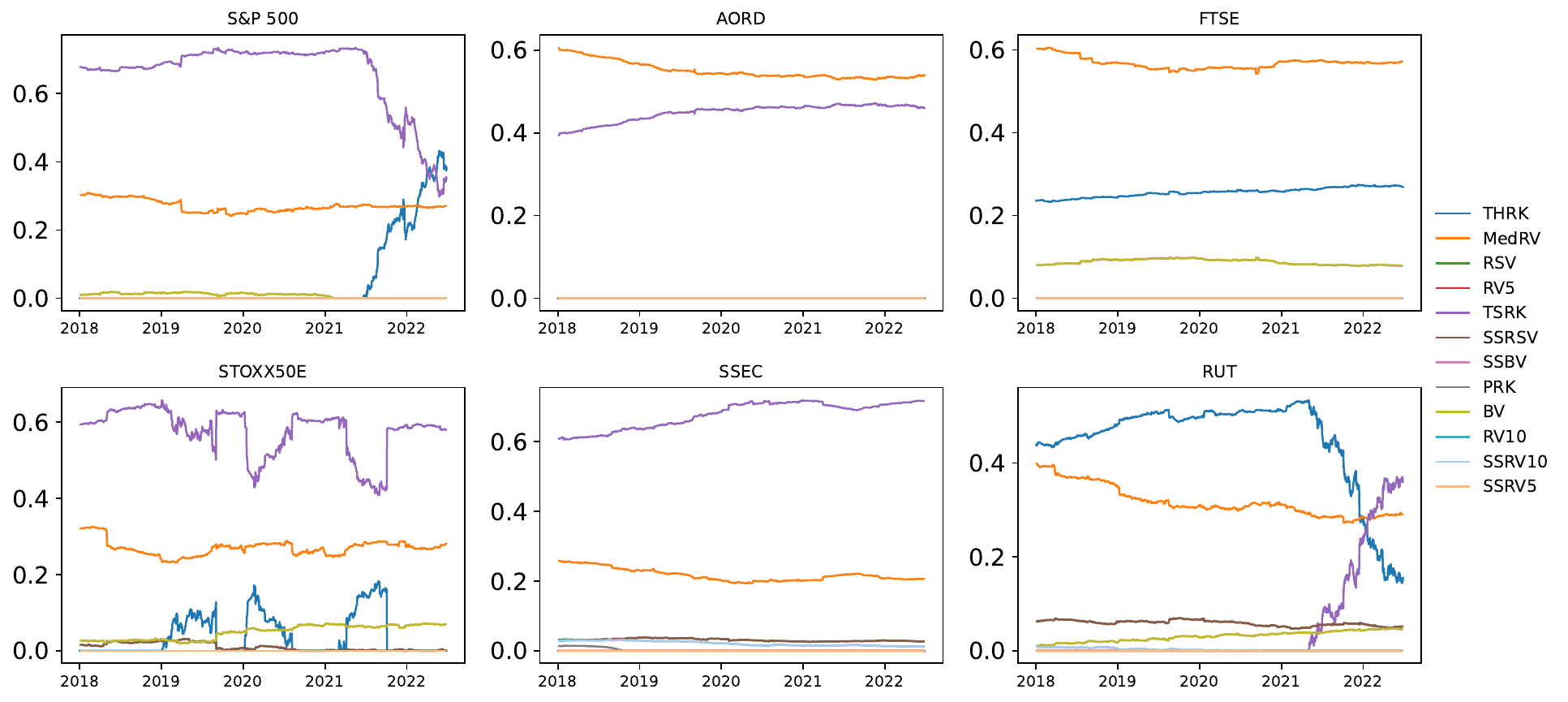}
    \caption[Rolling-window estimates of out-of-sample LW-RealGARCH weights.]{Rolling-window estimates of out-of-sample LW-RealGARCH weights, 2 January 2018--28 June 2022.}
    \label{fig:outsample_lw_weights}
\end{figure}

Figure~\ref{fig:outsample_lw_weights} presents the LW-RealGARCH weights estimated over the rolling out-of-sample windows. The weights vary substantially over time and are unevenly distributed across the candidate realized measures. In most windows, the weight is concentrated on a small subset of measures, particularly THRK, MedRV, and TSRK, whereas several other measures receive weights close to zero. Thus, the rolling estimation procedure reallocates the relative importance of the realized measures as the estimation window advances.

The dominant measures also change over time. For the S\&P 500, the initially high weight on TSRK declines after mid-2021, while the weight on THRK increases. The RUT displays the opposite pattern: the weight on THRK declines over time, whereas the weight on TSRK rises. The adjustment is particularly pronounced for STOXX50E at the beginning of 2020, when the TSRK weight falls sharply and the THRK weight increases during the COVID market disruption. These changes provide context for the COVID period forecast comparison in Section~\ref{sec:outsample_fc_pfm}.

\begin{figure}[H]
    \centering
    \includegraphics[width=\linewidth]{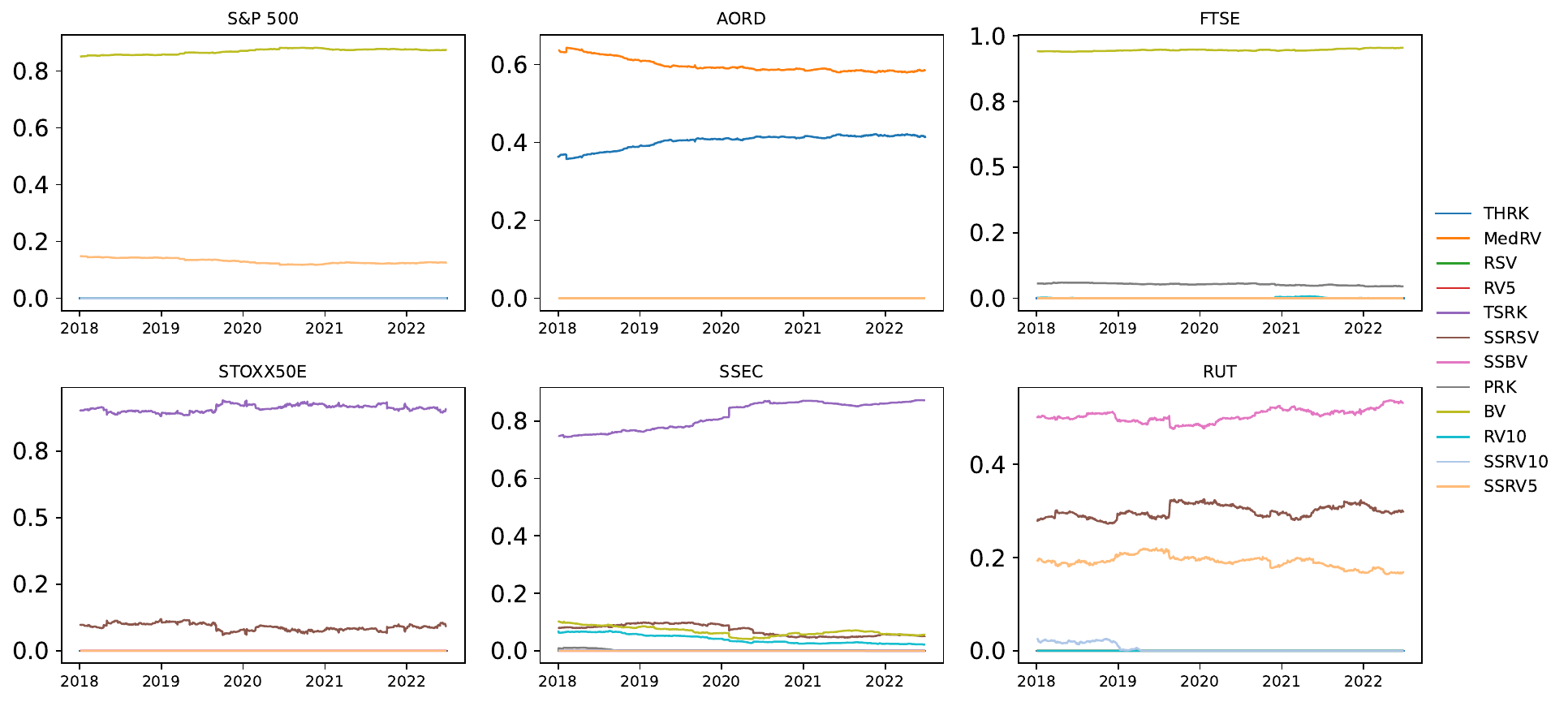}
    \caption[Rolling-window estimates of out-of-sample FW-RealGARCH weights.]{Rolling-window estimates of out-of-sample FW-RealGARCH weights, 2 January 2018--28 June 2022.}
    \label{fig:outsample_nlw_weights}
\end{figure}

Figure~\ref{fig:outsample_nlw_weights} presents the corresponding rolling-window weights from FW-RealGARCH. Relative to LW-RealGARCH, the FW-RealGARCH weights display more persistent patterns over time. This difference reflects the structure of the mixture of Beta weighting functions: although the estimated profile can change across windows, the individual weights remain linked through a common two-component weight profile. Consequently, FW-RealGARCH adjusts the aggregate measure mainly through changes in the estimated weight profile, whereas LW-RealGARCH can reallocate weight more freely across individual realized measures.
 
For the S\&P 500 and FTSE, BV receives relatively stable weights and remains dominant throughout the out-of-sample period. TSRK receives the greatest weight for STOXX50E and SSEC, with its weight increasing gradually in SSEC. For RUT, SSBV and SSRSV receive the largest weights, followed by SSRV5. AORD also displays a relatively stable profile: MedRV and THRK account for nearly all the weight throughout the period, with only modest changes in their relative contributions. Overall, the FW-RealGARCH profiles retain the same dominant measures across rolling estimation windows, with smaller adjustments than those observed under LW-RealGARCH.

\subsubsection{Forecasting Performance Evaluation}\label{sec:outsample_fc_pfm}
This section evaluates the out-of-sample forecasting performance of the competing models using the quantile score, AL joint loss, and NPLL introduced in Section~\ref{sec:forecasting_setup}. The three criteria assess VaR accuracy, joint VaR--ES accuracy, and volatility forecasts, respectively. Each table reports results for both the whole period and the COVID period. Lower values indicate better forecasting performance. We also report the model confidence set (MCS) at the 75\% confidence level to identify models that are statistically indistinguishable from the loss-minimizing specification \citep{hansen2011}.

Tables~\ref{tab:qloss_avg_combined}, \ref{tab:jointloss_combined}, and~\ref{tab:pnll_combined_1dp} report the quantile score, AL joint loss, and NPLL results, respectively. In general, the results provide broad evidence in favor of the proposed models. Under the quantile score, proposed models achieve the lowest loss for 23 of the 29 markets in both the whole period and COVID period panels, with LW-RealGARCH being the specification that most frequently generates the lowest quantile scores. A similar pattern is observed under the AL joint loss, where the proposed models attain the lowest loss for 23 markets over the whole period and 24 markets during the COVID period. For NPLL, the advantage is particularly clear during the COVID period, where the proposed models achieve the lowest NPLL for 26 markets. The main exception is FW-RealGARCH, which is generally less frequently selected as the best-performing model than LW-RealGARCH and the two AE-RealGARCH specifications. This result is consistent with the weight dynamics discussed in Section~\ref{sec:dynamic_weight_behavior}, where the parametric Beta-mixture weight profile restricts individual weight adjustments within each forecasting window.

The comparison between LW-RealGARCH and AVG-RealGARCH highlights the value of estimating market-specific weights rather than imposing equal weights across realized measures. Under the quantile score, LW-RealGARCH achieves the lowest loss for 13 markets in both the whole period and the COVID period, compared with 5 and 4 markets for AVG-RealGARCH, respectively. A similar pattern appears under AL joint loss, where LW-RealGARCH attains the lowest loss for 11 markets over the whole period and 7 markets during the COVID period, while AVG-RealGARCH does so for only 3 and 2 markets, respectively. For NPLL, the difference is less pronounced over the whole period, but it becomes clearer during the COVID period. LW-RealGARCH records 6 lowest loss cases and 23 MCS inclusions, compared with 2 and 20 for AVG-RealGARCH. These results suggest that adaptive weighting provides a clear forecasting gain over simple averaging, especially when market conditions become more volatile.

\begin{figure}[htbp]
    \centering
    \includegraphics[width=0.8\linewidth]{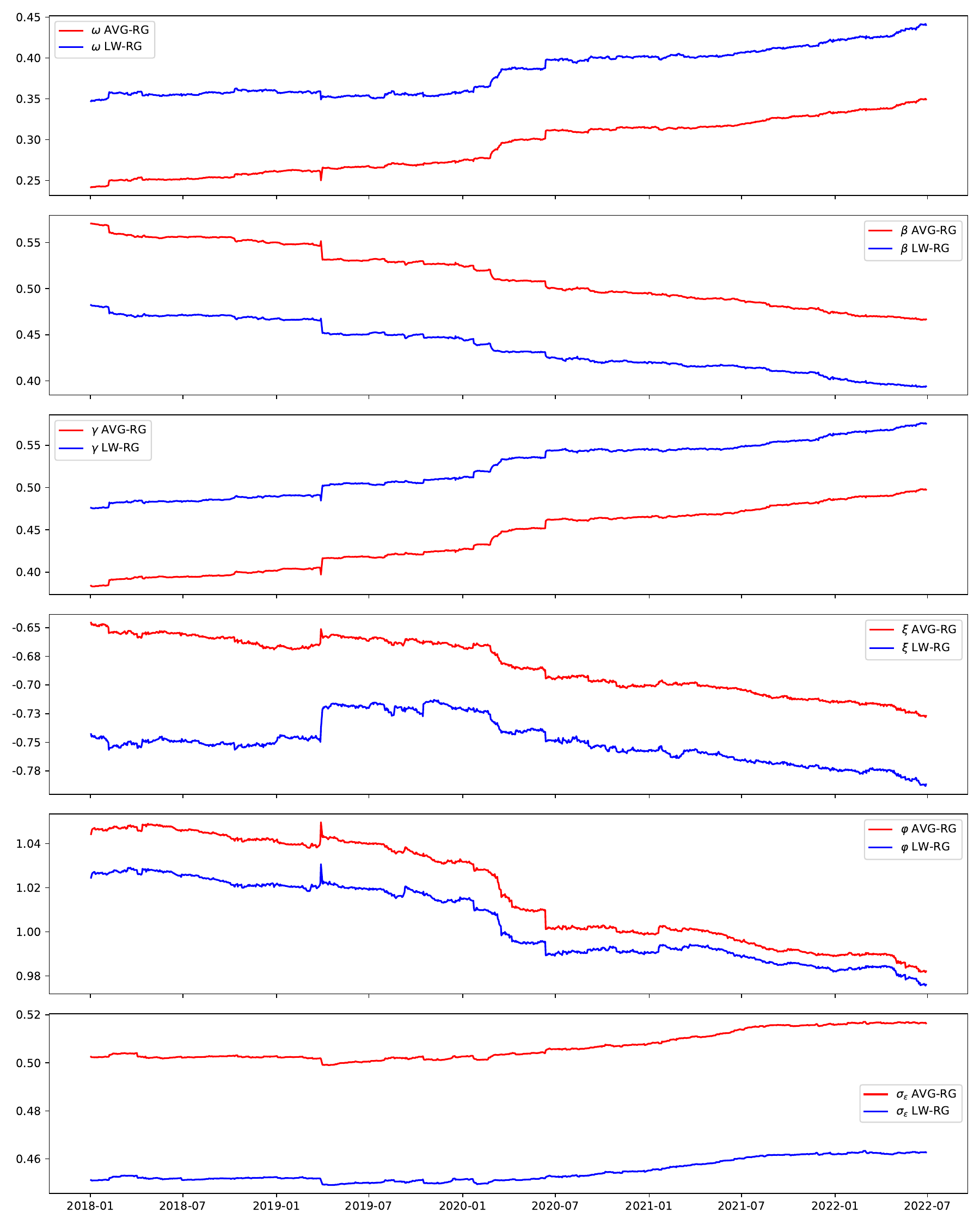}
    \caption{Rolling out-of-sample parameter estimates for the S\&P 500 under LW-RealGARCH and AVG-RealGARCH.}
    \label{fig:outofsample_lw_avg_compare}
\end{figure}

To further illustrate why LW-RealGARCH delivers stronger forecasting performance than AVG-RealGARCH, Figure~\ref{fig:outofsample_lw_avg_compare} compares the out-of-sample S\&P 500 parameter estimates under the two specifications. Overall, the two models exhibit similar time-series patterns, but their differences are most evident in $\gamma$, which measures the importance assigned to the lagged synthetic realized measure. In particular, the estimated $\gamma$ coefficient is consistently higher for LW-RealGARCH than for AVG-RealGARCH throughout the out-of-sample period, indicating that the linear weighted measure provides a stronger volatility signal than the simple average measure. Correspondingly, the estimated $\beta$ coefficient, which captures the effect of lagged conditional volatility, is consistently lower for LW-RealGARCH. The measurement-equation parameter $\varphi$ remains close to unity for both models, and the measurement-error standard deviation $\sigma_\varepsilon$ is lower for LW-RealGARCH, suggesting that the linear weighted measure is more closely aligned with the latent volatility process than the equally weighted average measure. Taken together, these parameter patterns indicate that the forecasting gain of LW-RealGARCH primarily stems from a more precise realized measure while preserving the core Realized GARCH dynamics.

Within the two-step dimension-reduction specifications, the two AE-RealGARCH models also perform favorably relative to PC-RealGARCH and IC-RealGARCH. Under the quantile score, the two AE-RealGARCH models jointly record 8 lowest loss cases over the whole period and 8 during the COVID period, whereas PC-RealGARCH and IC-RealGARCH record no cases in either panel. During the COVID period, AE-RealGARCH with the 12-12-1 architecture is included in the quantile score MCS for 28 of the 29 markets, exceeding both PC-RealGARCH and IC-RealGARCH by 3 markets. Under the AL joint loss, the two AE-RealGARCH specifications jointly attain 11 lowest loss cases over the whole period and 12 during the COVID period, while PC-RealGARCH and IC-RealGARCH again record none. During the COVID period, the 12-6-1 and 12-12-1 AE-RealGARCH models are retained in the AL joint loss MCS for 24 and 26 markets, respectively, exceeding both PC-RealGARCH and IC-RealGARCH benchmarks by 3 and 5 markets. For NPLL, the AE-RealGARCH models record 5 lowest loss cases over the whole period and 15 during the COVID period, compared with 1 and 0 cases, respectively, for PC-RealGARCH and IC-RealGARCH combined. Overall, these results indicate that the nonlinear autoencoder method provides more informative synthetic realized measures than the linear PCA and ICA transformations, with the advantage becoming especially clear during the COVID period.

\input{Table/qloss_avg_combined_table}

\input{Table/jointloss_combined_table}

\input{Table/pnll_combined_table_1dp}

\section{Conclusion}\label{sec:conclusion}
This project examines how multiple realized measures can be incorporated into the Realized GARCH framework. We propose three Realized GARCH extensions: LW-RealGARCH and FW-RealGARCH construct synthetic realized measures through data-driven weighting schemes, while AE-RealGARCH uses an autoencoder to summarize common nonlinear dynamics across realized-measure time series. All three models preserve the standard Realized GARCH structure while incorporating information from a broader set of realized measures. Based on 29 international equity indices, the empirical results support the proposed models. LW-RealGARCH and FW-RealGARCH identify market-specific weighting patterns, highlighting the importance of adaptive realized-measure aggregation. The AE-RealGARCH results show that the autoencoder produces an informative synthetic measure, with particularly useful performance during the COVID period, when financial risk is elevated.

Future research could extend the functional weighting component of the proposed framework. Although FW-RealGARCH uses a parsimonious mixture of Beta weighting profiles, alternative nonlinear weighting schemes, such as the relative-score combination approach of \citet{taylor2020a}, could allow aggregation weights to respond more directly to forecasting performance while retaining a clear weighting interpretation.

Another direction is to significantly expand the information set used by the autoencoder. Because of data availability, this study uses 12 realized measures, but the proposed architecture is ready to be extended to much higher-dimensional inputs, including hundreds of candidate variables, with appropriate regularization. Future applications could also test incorporating uncertainty indices and news-based indicators, such as the economic policy uncertainty index of \citet{baker2016} and the equity market volatility index of \citet{baker2019}. These variables may contain macroeconomic and news-based uncertainty information not fully captured by realized volatility measures alone, allowing the autoencoder to learn a broader representation of financial volatility and risk to further improve their forecasting performance.
\section*{Funding}
This research did not receive any specific grant from funding agencies in the public, commercial, or not-for-profit sectors.

\section*{Data availability}
This study uses data from the Oxford-Man Institute's Realised Library \citep{gerdheber2009}.
The data are available at \url{https://github.com/...}\footnote{Github link anonymized for double anonymized peer review.}.
\printbibliography

\clearpage
\appendix

\section{Autoencoder Training Configurations}
\label{app:ae_training_config}

\setcounter{figure}{0}
\setcounter{table}{0}

\renewcommand{\thefigure}{A\arabic{figure}}
\renewcommand{\thetable}{A\arabic{table}}

\renewcommand{\theHfigure}{appendix.A.\arabic{figure}}
\renewcommand{\theHtable}{appendix.A.\arabic{table}}

To ensure a consistent and reproducible construction of the autoencoder-based realized measure, we use the same configuration across the two AE-RealGARCH architectures. The configuration was selected after extensive tuning, with the aim of obtaining a stable specification that can be generalized across all markets. Since the autoencoder is trained in an unsupervised manner, these settings are designed to ensure that the network produces a stable and reasonable nonlinear compression of the realized measures, while remaining suitable for time-series applications.

The loss function, weight regularization term, sparsity penalty, and adaptive target activation are defined in Section~\ref{sec:ae_training}. This appendix reports the numerical settings used in the implementation as specified in Table \ref{tab:ae_training_config}. The autoencoder is implemented in PyTorch, which also ensures reproducibility after the future removal of the built-in autoencoder functionality from MATLAB. The complete code used to train the autoencoder on time-series realized measures is available in the accompanying GitHub repository: \url{https://github.com/...}\footnote{Github link anonymized for double anonymized peer review.}. The autoencoder was trained on a consumer grade with an NVIDIA GeForce RTX 5060 Ti GPU and an Intel Core i5-14600K CPU. The autoencoder parameters are trained by backpropagation. At each epoch, gradients of the total loss with respect to all trainable weights and biases are computed through backpropagation, and the parameters are then updated using the Adam optimizer \citep{kingma2017}.

\begin{table}[htbp]
\centering
\caption{Autoencoder training configurations}
\label{tab:ae_training_config}
\small
\setlength{\tabcolsep}{5pt}
\renewcommand{\arraystretch}{1.15}
\begin{tabular}{p{0.25\textwidth}p{0.18\textwidth}>{\raggedright\arraybackslash}p{0.48\textwidth}}
\hline
Component & Setting & Description \\
\hline
Optimizer & Adam & Used with backpropagation to update all trainable autoencoder parameters. \\
Learning rate & $0.1$ & Fixed learning rate used during training. \\
Maximum epochs & $200$ & Maximum number of epochs. \\
Weight regularization & $\lambda_1=0.1$ & Implemented through Adam weight decay. \\
Sparsity weights & $\lambda_2=1.0$ & KL-divergence weights for the two encoder layers. \\
Early stopping patience & $10$ & Number of epochs allowed without sufficient improvement. \\
Minimum improvement & $10^{-2}$ & Minimum decrease in total loss required to reset patience. \\
\hline
\end{tabular}
\end{table}

Each run uses the full training tensor as the input batch. At each epoch, the autoencoder reconstructs the scaled realized measure inputs, and the total loss is computed as the sum of the reconstruction loss, weight penalty, and the sparsity penalty. Early stopping is based on the total loss. If the loss decreases by more than $10^{-2}$ relative to the best previous value, the current parameter state is stored and the patience counter is reset. Otherwise, the patience counter increases by one. Training stops when the counter reaches 10, or when the maximum of 200 epochs is reached. After training, the model is reloaded using the parameter state associated with the lowest total loss.

\begin{figure}[htbp]
    \centering
    \includegraphics[width=\linewidth]{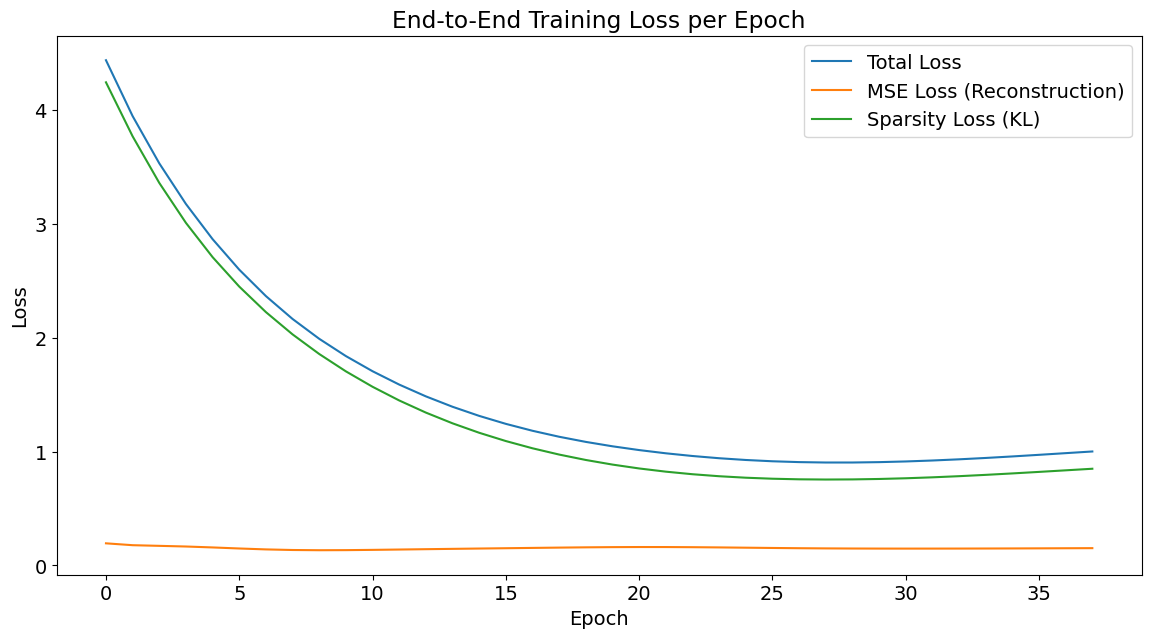}
    \caption{Autoencoder training loss for the in-sample S\&P 500 data.}
    \label{fig:ae_training_loss}
\end{figure}

Figure~\ref{fig:ae_training_loss} illustrates the autoencoder training loss for the in-sample S\&P 500 data. The total loss decreases rapidly during the early epochs and then stabilizes, indicating that the selected configuration achieves convergence within a relatively small number of iterations. The total loss closely follows the sparsity loss, showing that the KL-divergence sparsity penalty is the dominant component of the training objective and plays a key role in shaping the learned representation. The MSE reconstruction loss remains comparatively small and stable throughout training. The weight regularization component corresponds to the difference between the total loss and the sum of the MSE reconstruction loss and sparsity loss. The early-stopping rule further prevents unnecessary additional updates after the loss has stabilized, reducing the risk of overfitting while preserving a stable autoencoder-based realized measure.

\begin{figure}
    \centering
    \includegraphics[
        width=\linewidth,
        height=0.9\textheight,
        keepaspectratio
    ]{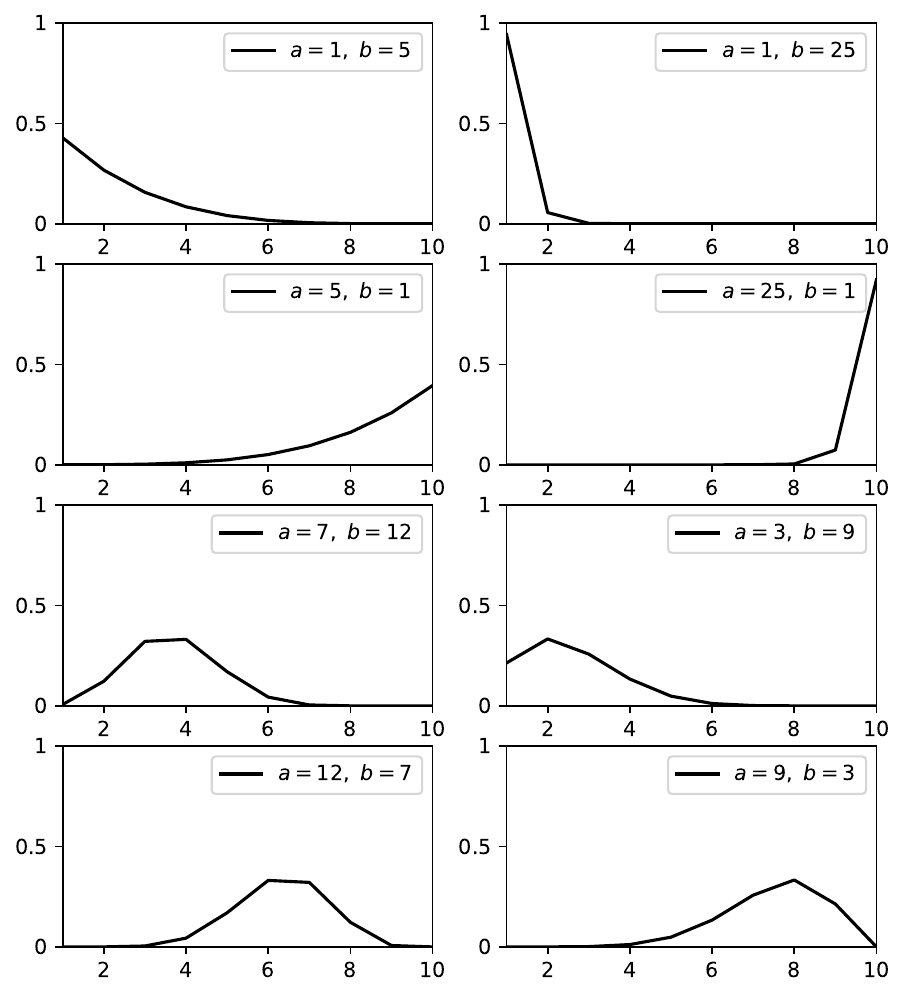}
    \caption{Beta weight profiles for different parameter values.}
    \label{fig:appendix_beta_weights}
\end{figure}

\begin{figure}
    \centering
    \includegraphics[width=1\linewidth]{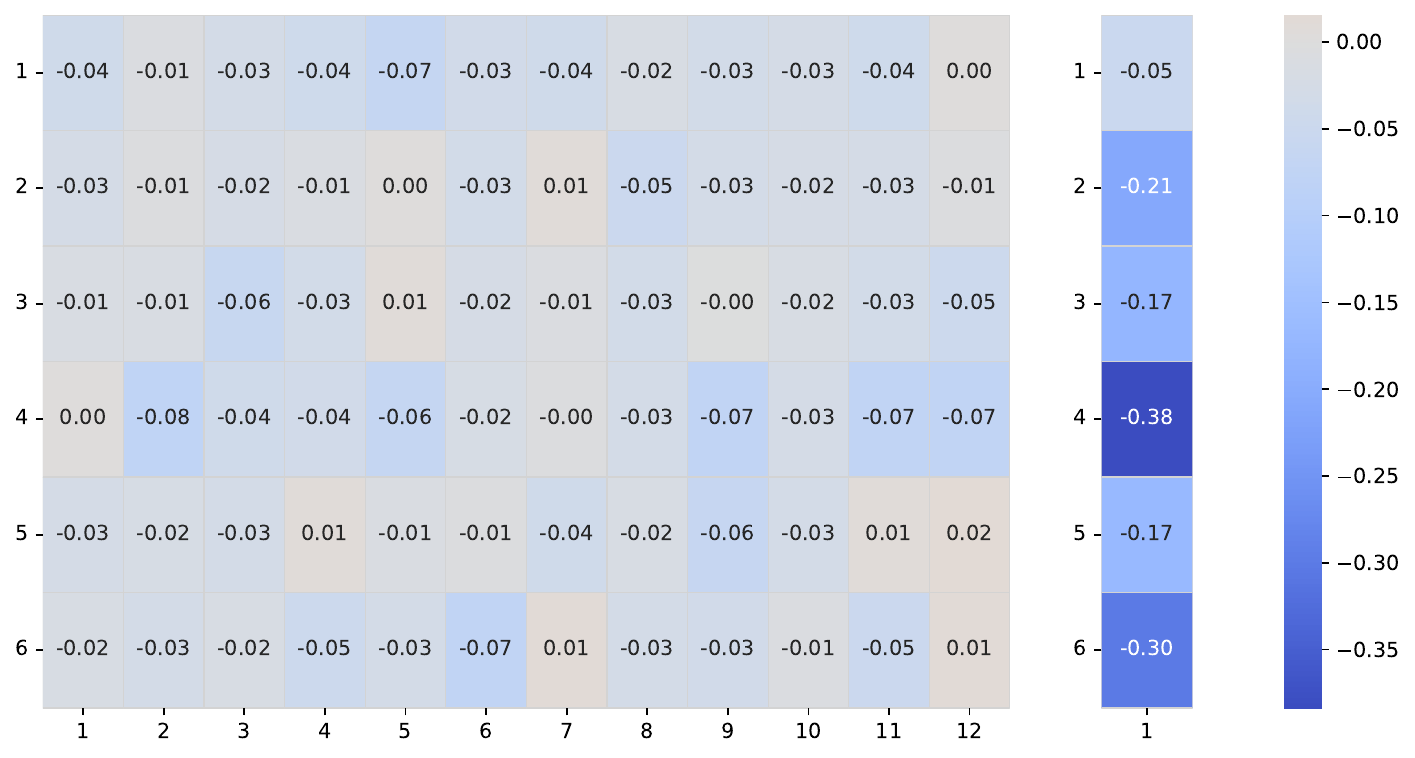}
    \caption{Heatmaps of the encoder weight matrices for the 12-6-1-6-12 autoencoder. The left panel shows the weights connecting the 12-dimensional input layer to the six-unit hidden layer, while the right panel shows the weights connecting the hidden layer to the one-unit bottleneck layer.}
    \label{fig:appendix_encoder_wts_hm}
\end{figure}

\end{document}

%% file: Table/rt_stats.tex
\begin{table}[H]
    \centering
    \small
    \caption{Descriptive Statistics for Return series of 29 Markets.}
    \label{tab:rt_stats}
    \begin{tabular}{lccccccc}
    \toprule
    \textbf{Market} & \textbf{T} & \textbf{Mean} & \textbf{Std Dev} & \textbf{Min} & \textbf{Max} & \textbf{Skewness} & \textbf{Kurtosis} \\
    \midrule
    AEX & 5731 & -0.0004 & 1.3736 & -10.6385 & 9.5718 & -0.2648 & 6.8242 \\
    AORD & 5668 & 0.0139 & 0.9427 & -7.2610 & 4.5327 & -0.7597 & 6.1863 \\
    BFX & 5728 & 0.0020 & 1.2459 & -14.2213 & 9.3028 & -0.3655 & 8.8807 \\
    BSESN & 5564 & 0.0425 & 1.4428 & -13.7811 & 16.1147 & -0.2945 & 8.6406 \\
    BVSP & 5528 & 0.0327 & 1.7935 & -15.9938 & 13.8163 & -0.3081 & 6.7826 \\
    DJI & 5626 & 0.0180 & 1.1870 & -13.8069 & 10.7538 & -0.3966 & 12.7172 \\
    FCHI & 5735 & 0.0004 & 1.4124 & -11.9972 & 10.4387 & -0.2299 & 5.8079 \\
    FTMIB & 3314 & 0.0023 & 1.5652 & -18.5411 & 10.6840 & -0.8963 & 10.4818 \\
    FTSE & 5662 & 0.0012 & 1.1620 & -10.1365 & 9.4849 & -0.3061 & 7.1209 \\
    GDAXI & 5694 & 0.0114 & 1.4567 & -11.8631 & 12.0269 & -0.1569 & 6.2955 \\
    GSPTSE & 5035 & 0.0183 & 1.0448 & -13.1761 & 9.1202 & -1.1842 & 16.5861 \\
    HSI & 5504 & 0.0040 & 1.4673 & -13.5820 & 13.4068 & -0.0812 & 7.4245 \\
    IBEX & 5696 & -0.0066 & 1.4471 & -12.7177 & 12.8716 & -0.3161 & 7.0519 \\
    IXIC & 5635 & 0.0177 & 1.5928 & -13.1409 & 13.2782 & -0.1866 & 6.3701 \\
    KS11 & 5527 & 0.0150 & 1.4599 & -12.8250 & 11.2449 & -0.6164 & 7.5976 \\
    KSE & 5426 & 0.0678 & 1.3385 & -8.4636 & 8.5071 & -0.4014 & 4.3396 \\
    MXX & 5633 & 0.0354 & 1.2604 & -8.2673 & 10.4407 & -0.0525 & 5.3525 \\
    N225 & 5459 & 0.0059 & 1.4765 & -12.1110 & 13.2346 & -0.3830 & 6.1565 \\
    NSEI & 5239 & 0.0510 & 1.3976 & -13.6329 & 16.2255 & -0.3219 & 10.0903 \\
    OMXC20 & 4149 & 0.0354 & 1.2924 & -11.7232 & 9.4964 & -0.3117 & 5.6436 \\
    OMXHPI & 4194 & 0.0069 & 1.3365 & -10.7876 & 8.8500 & -0.2899 & 5.2732 \\
    OMXSPI & 4191 & 0.0240 & 1.3237 & -11.8051 & 10.0345 & -0.3297 & 7.0948 \\
    OSEAX & 5181 & 0.0406 & 1.3654 & -9.8320 & 9.1864 & -0.6860 & 6.7842 \\
    RUT & 5630 & 0.0225 & 1.5610 & -15.2290 & 8.9056 & -0.5271 & 6.8796 \\
    SMSI & 4320 & -0.0060 & 1.4190 & -14.0609 & 12.0710 & -0.4608 & 8.8461 \\
    S\&P 500 & 5634 & 0.0172 & 1.2420 & -12.6703 & 10.6420 & -0.3930 & 10.2096 \\
    SSEC & 5422 & 0.0175 & 1.5106 & -9.2114 & 9.5055 & -0.4021 & 5.0084 \\
    SSMI & 5632 & 0.0074 & 1.1539 & -10.1339 & 10.7876 & -0.3259 & 8.2367 \\
    STOXX50E & 5716 & -0.0053 & 1.4307 & -12.0054 & 10.5536 & -0.2058 & 5.6487 \\
    \bottomrule
    \end{tabular}
\end{table}

%% file: Table/spx_log_rm_stats.tex
\begin{table}[htbp]
\centering
\caption{Descriptive statistics of in-sample log-transformed realized measures for S\&P 500.}
\label{tab:spx_log_rm_stats}
\small
\begin{tabular}{lccccccc}
\toprule
 & Mean & Std. & Median & Min & Max & Skewness & Kurtosis \\
\midrule
$\log(x_{\mathrm{AE}})$ (12-6-1)   & -0.418 & 1.179 & -0.454 & -6.101 & 4.444 & 0.186 & 0.471 \\
$\log(x_{\mathrm{AE}})$ (12-12-1)  & -0.456 & 1.176 & -0.496 & -6.101 & 4.444 & 0.193 & 0.470 \\
$\log(x_{\mathrm{LW}})$            & -0.939 & 1.105 & -1.006 & -4.398 & 4.196 & 0.382 & 0.426 \\
$\log(x_{\mathrm{FW}})$           & -0.822 & 1.118 & -0.880 & -4.474 & 4.139 & 0.371 & 0.407 \\
$\log(x_{\mathrm{AVG}})$           & -0.845 & 1.129 & -0.903 & -4.524 & 4.135 & 0.340 & 0.359 \\
$\log(x_{\mathrm{PC}})$            & -0.582 & 1.171 & -0.617 & -6.101 & 4.444 & 0.195 & 0.496 \\
$\log(x_{\mathrm{IC}})$            & -0.582 & 1.171 & -0.617 & -6.101 & 4.444 & 0.195 & 0.496 \\
$\log(x_{\mathrm{RV5}})$           & -0.643 & 1.135 & -0.692 & -4.408 & 4.350 & 0.323 & 0.354 \\
\bottomrule
\end{tabular}
\caption*{\scriptsize{Note: $x_{\mathrm{AE}}$ denotes the synthetic realized measure constructed using the autoencoder method, with the two rows corresponding to the 12-6-1 and 12-12-1 autoencoder architectures, respectively. $x_{\mathrm{LW}}$, $x_{\mathrm{FW}}$, $x_{\mathrm{AVG}}$, $x_{\mathrm{PC}}$, and $x_{\mathrm{IC}}$ denote synthetic realized measures constructed using the linear weighting, functional weighting, averaging, PCA, and ICA methods, respectively. $x_{\mathrm{RV5}}$ denotes the traditional 5-minute realized variance measure.}}

\end{table}

%% file: Table/insample_params_v2.tex
\begin{table}[htbp]
\centering
\caption{In-sample estimated parameters for RealGARCH framework models in S\&P 500.}
\label{tab:insample_params_spx}
\small
\setlength{\tabcolsep}{3pt}
\resizebox{\textwidth}{!}{%
\begin{tabular}{lcccc:cccc}
\hline
& \multicolumn{4}{c:}{Proposed models}
& \multicolumn{4}{c}{Benchmarks} \\
& \begin{tabular}{@{}c@{}}AE-RG\\(12-6-1)\end{tabular}
& \begin{tabular}{@{}c@{}}AE-RG\\(12-12-1)\end{tabular}
& LW-RG
& {\setbox0=\hbox{NLW-RG}\makebox[\wd0][c]{FW-RG}}
& AVG-RG
& PC-RG
& IC-RG
& RG \\
\hline
$\omega$     & 0.073 & 0.087 & 0.347 & 0.244 & 0.242 & 0.133 & 0.133 & 0.154 \\
$\beta$      & 0.572 & 0.574 & 0.482 & 0.557 & 0.571 & 0.574 & 0.574 & 0.598 \\
$\gamma$     & 0.372 & 0.370 & 0.476 & 0.404 & 0.384 & 0.367 & 0.367 & 0.357 \\
$\xi$        & -0.215 & -0.252 & -0.743 & -0.622 & -0.646 & -0.380 & -0.380 & -0.447 \\
$\varphi$       & 1.074 & 1.074 & 1.024 & 1.028 & 1.044 & 1.083 & 1.083 & 1.049 \\
$\tau_1$     & -0.187 & -0.183 & -0.147 & -0.142 & -0.145 & -0.133 & -0.133 & -0.101 \\
$\tau_2$     & 0.106 & 0.107 & 0.071 & 0.080 & 0.106 & 0.114 & 0.114 & 0.116 \\
$\sigma_\varepsilon$ & 0.530 & 0.528 & 0.451 & 0.496 & 0.503 & 0.532 & 0.532 & 0.537 \\
$\omega+\gamma\xi$ & -0.0063 & -0.0063 & -0.0066 & -0.0071 & -0.0062 & -0.0061 & -0.0061 & -0.0060 \\
NLL          & 3671.0 & 3672.0 & 3663.3 & 3644.5 & 3665.9 & 3687.6 & 3687.6 & 3683.9 \\
Time         & $0.14+0.11$ & $0.14+0.12$ & 0.39 & 0.14 & $0.00+0.12$ & $0.03+0.10$ & $0.04+0.11$ & 0.12 \\
\hline
\end{tabular}%
}
\caption*{\scriptsize{Note: All reported values are averages over 10 repeated trials to reduce randomness in optimization. AE-RG (12-6-1) and AE-RG (12-12-1) denote AE-RealGARCH models using autoencoder architectures of 12-6-1 and 12-12-1, respectively. LW-RG, FW-RG, AVG-RG, PC-RG, and IC-RG denote LW-RealGARCH, FW-RealGARCH, AVG-RealGARCH, PC-RealGARCH, and IC-RealGARCH, respectively. RG denotes the RealGARCH model using RV5 as the realized measure. NLL denotes the in-sample GARCH negative log-likelihood.}}

\end{table}

%% file: Table/qloss_avg_combined_table.tex
\clearpage
\begin{table}[htbp]
\centering
\caption{Average quantile score comparison and MCS results.}
\label{tab:qloss_avg_combined}
\tiny
\renewcommand{\arraystretch}{0.8}
\newcolumntype{C}{>{\centering\arraybackslash}p{4.5em}}
\begin{tabular}{lCCCC:CCCC}
\hline
Whole period & \multicolumn{4}{c:}{Proposed Models}
& \multicolumn{4}{c}{Benchmarks} \\
01/01/2018--28/06/2022 & \begin{tabular}{@{}c@{}}AE-RG\\(12-6-1)\end{tabular} & \begin{tabular}{@{}c@{}}AE-RG\\(12-12-1)\end{tabular} & LW-RG & FW-RG & AVG-RG & PC-RG & IC-RG & RG \\
\hline
S\&P 500 & \cellcolor{gray!25} 0.0868 & \cellcolor{gray!25} 0.0869 & \cellcolor{gray!25} 0.0850 & \cellcolor{gray!25} \textcolor{blue}{0.0848} & 0.0871 & 0.0877 & 0.0877 & 0.0892 \\
IXIC & \cellcolor{gray!25} 0.0939 & \cellcolor{gray!25} 0.0939 & \cellcolor{gray!25} 0.0971 & \cellcolor{gray!25} 0.0966 & \cellcolor{gray!25} \textcolor{blue}{0.0938} & \cellcolor{gray!25} 0.0944 & \cellcolor{gray!25} 0.0944 & \cellcolor{gray!25} 0.0940 \\
DJI & 0.0896 & 0.0898 & \cellcolor{gray!25} \textcolor{blue}{0.0851} & 0.0873 & 0.0897 & 0.0903 & 0.0903 & 0.0923 \\
SSEC & \cellcolor{gray!25} 0.0867 & \cellcolor{gray!25} 0.0865 & \cellcolor{gray!25} \textcolor{blue}{0.0853} & 0.0862 & \cellcolor{gray!25} 0.0865 & 0.0877 & 0.0877 & \cellcolor{gray!25} 0.0871 \\
N225 & 0.0848 & 0.0847 & \cellcolor{gray!25} \textcolor{blue}{0.0839} & 0.0860 & 0.0851 & 0.0855 & 0.0855 & 0.0859 \\
HSI & \cellcolor{gray!25} 0.0925 & \cellcolor{gray!25} 0.0923 & \cellcolor{gray!25} \textcolor{blue}{0.0923} & \cellcolor{gray!25} 0.0930 & \cellcolor{gray!25} 0.0928 & \cellcolor{gray!25} 0.0926 & \cellcolor{gray!25} 0.0926 & \cellcolor{gray!25} 0.0931 \\
STOXX50E & \cellcolor{gray!25} \textcolor{blue}{0.0890} & 0.0894 & 0.0914 & 0.0927 & 0.0914 & 0.0917 & 0.0917 & 0.0923 \\
NSEI & \cellcolor{gray!25} 0.0838 & \cellcolor{gray!25} 0.0844 & \cellcolor{gray!25} \textcolor{blue}{0.0836} & \cellcolor{gray!25} 0.0844 & \cellcolor{gray!25} 0.0839 & \cellcolor{gray!25} 0.0844 & \cellcolor{gray!25} 0.0844 & \cellcolor{gray!25} 0.0845 \\
FTSE & 0.0815 & \cellcolor{gray!25} \textcolor{blue}{0.0802} & 0.0845 & 0.0838 & 0.0849 & 0.0852 & 0.0852 & 0.0852 \\
BSESN & \cellcolor{gray!25} 0.0829 & \cellcolor{gray!25} 0.0829 & \cellcolor{gray!25} 0.0830 & \cellcolor{gray!25} 0.0836 & \cellcolor{gray!25} \textcolor{blue}{0.0821} & \cellcolor{gray!25} 0.0826 & \cellcolor{gray!25} 0.0826 & \cellcolor{gray!25} 0.0825 \\
GSPTSE & \cellcolor{gray!25} 0.0729 & \cellcolor{gray!25} 0.0722 & \cellcolor{gray!25} \textcolor{blue}{0.0684} & 0.0697 & \cellcolor{gray!25} 0.0703 & \cellcolor{gray!25} 0.0713 & \cellcolor{gray!25} 0.0713 & 0.0728 \\
FCHI & \cellcolor{gray!25} 0.0925 & \cellcolor{gray!25} 0.0925 & \cellcolor{gray!25} \textcolor{blue}{0.0920} & \cellcolor{gray!25} 0.0928 & \cellcolor{gray!25} 0.0924 & 0.0928 & 0.0928 & \cellcolor{gray!25} 0.0927 \\
RUT & \cellcolor{gray!25} 0.1035 & \cellcolor{gray!25} 0.1037 & \cellcolor{gray!25} 0.1043 & \cellcolor{gray!25} 0.1035 & \cellcolor{gray!25} \textcolor{blue}{0.1029} & \cellcolor{gray!25} 0.1039 & \cellcolor{gray!25} 0.1039 & \cellcolor{gray!25} 0.1042 \\
GDAXI & 0.0986 & \cellcolor{gray!25} 0.0984 & \cellcolor{gray!25} 0.0988 & \cellcolor{gray!25} \textcolor{blue}{0.0981} & 0.0988 & 0.0989 & 0.0989 & 0.0995 \\
KS11 & \cellcolor{gray!25} 0.0729 & \cellcolor{gray!25} 0.0728 & \cellcolor{gray!25} \textcolor{blue}{0.0725} & \cellcolor{gray!25} 0.0726 & \cellcolor{gray!25} 0.0728 & \cellcolor{gray!25} 0.0730 & \cellcolor{gray!25} 0.0730 & \cellcolor{gray!25} 0.0736 \\
SSMI & \cellcolor{gray!25} \textcolor{blue}{0.0697} & \cellcolor{gray!25} 0.0697 & 0.0726 & \cellcolor{gray!25} 0.0700 & \cellcolor{gray!25} 0.0700 & \cellcolor{gray!25} 0.0701 & \cellcolor{gray!25} 0.0701 & \cellcolor{gray!25} 0.0700 \\
AORD & \cellcolor{gray!25} \textcolor{blue}{0.0741} & \cellcolor{gray!25} 0.0741 & \cellcolor{gray!25} 0.0746 & \cellcolor{gray!25} 0.0748 & 0.0748 & \cellcolor{gray!25} 0.0750 & \cellcolor{gray!25} 0.0750 & 0.0751 \\
BVSP & \cellcolor{gray!25} 0.1173 & \cellcolor{gray!25} 0.1172 & \cellcolor{gray!25} \textcolor{blue}{0.1137} & \cellcolor{gray!25} 0.1159 & \cellcolor{gray!25} 0.1148 & \cellcolor{gray!25} 0.1173 & \cellcolor{gray!25} 0.1173 & \cellcolor{gray!25} 0.1151 \\
IBEX & \cellcolor{gray!25} 0.0904 & \cellcolor{gray!25} 0.0903 & \cellcolor{gray!25} 0.0904 & \cellcolor{gray!25} 0.0906 & \cellcolor{gray!25} \textcolor{blue}{0.0902} & \cellcolor{gray!25} 0.0907 & \cellcolor{gray!25} 0.0907 & \cellcolor{gray!25} 0.0902 \\
FTMIB & \cellcolor{gray!25} 0.1088 & \cellcolor{gray!25} \textcolor{blue}{0.1088} & \cellcolor{gray!25} 0.1103 & \cellcolor{gray!25} 0.1090 & \cellcolor{gray!25} 0.1089 & \cellcolor{gray!25} 0.1089 & \cellcolor{gray!25} 0.1089 & \cellcolor{gray!25} 0.1091 \\
AEX & \cellcolor{gray!25} 0.0815 & \cellcolor{gray!25} \textcolor{blue}{0.0815} & \cellcolor{gray!25} 0.0821 & 0.0827 & 0.0821 & 0.0821 & 0.0821 & \cellcolor{gray!25} 0.0819 \\
OMXSPI & \cellcolor{gray!25} \textcolor{blue}{0.0838} & \cellcolor{gray!25} 0.0838 & \cellcolor{gray!25} 0.0854 & \cellcolor{gray!25} 0.0865 & \cellcolor{gray!25} 0.0866 & \cellcolor{gray!25} 0.0864 & \cellcolor{gray!25} 0.0864 & 0.0874 \\
MXX & 0.0757 & 0.0756 & 0.0763 & \cellcolor{gray!25} 0.0752 & \cellcolor{gray!25} \textcolor{blue}{0.0748} & \cellcolor{gray!25} 0.0749 & \cellcolor{gray!25} 0.0749 & \cellcolor{gray!25} 0.0754 \\
BFX & \cellcolor{gray!25} 0.0837 & \cellcolor{gray!25} 0.0837 & \cellcolor{gray!25} \textcolor{blue}{0.0830} & 0.0846 & \cellcolor{gray!25} 0.0835 & \cellcolor{gray!25} 0.0838 & \cellcolor{gray!25} 0.0838 & \cellcolor{gray!25} 0.0842 \\
OSEAX & \cellcolor{gray!25} 0.0805 & \cellcolor{gray!25} \textcolor{blue}{0.0804} & \cellcolor{gray!25} 0.0826 & \cellcolor{gray!25} 0.0816 & \cellcolor{gray!25} 0.0806 & \cellcolor{gray!25} 0.0811 & \cellcolor{gray!25} 0.0811 & \cellcolor{gray!25} 0.0813 \\
OMXC20 & \cellcolor{gray!25} 0.0812 & \cellcolor{gray!25} 0.0811 & \cellcolor{gray!25} \textcolor{blue}{0.0783} & \cellcolor{gray!25} 0.0790 & \cellcolor{gray!25} 0.0794 & \cellcolor{gray!25} 0.0800 & \cellcolor{gray!25} 0.0800 & 0.0801 \\
OMXHPI & \cellcolor{gray!25} 0.0831 & \cellcolor{gray!25} 0.0830 & \cellcolor{gray!25} \textcolor{blue}{0.0791} & \cellcolor{gray!25} 0.0799 & \cellcolor{gray!25} 0.0797 & \cellcolor{gray!25} 0.0801 & \cellcolor{gray!25} 0.0801 & 0.0806 \\
SMSI & \cellcolor{gray!25} 0.0867 & \cellcolor{gray!25} 0.0869 & \cellcolor{gray!25} \textcolor{blue}{0.0864} & \cellcolor{gray!25} 0.0865 & \cellcolor{gray!25} 0.0865 & \cellcolor{gray!25} 0.0871 & \cellcolor{gray!25} 0.0871 & \cellcolor{gray!25} 0.0866 \\
KSE & 0.0847 & 0.0847 & \cellcolor{gray!25} 0.0846 & \cellcolor{gray!25} 0.0846 & \cellcolor{gray!25} 0.0841 & 0.0847 & 0.0847 & \cellcolor{gray!25} \textcolor{blue}{0.0841} \\
\hline
COVID period & \multicolumn{4}{c:}{Proposed Models}
& \multicolumn{4}{c}{Benchmarks} \\
30/01/2020--30/06/2020 & \begin{tabular}{@{}c@{}}AE-RG\\(12-6-1)\end{tabular} & \begin{tabular}{@{}c@{}}AE-RG\\(12-12-1)\end{tabular} & LW-RG & FW-RG & AVG-RG & PC-RG & IC-RG & RG \\
\hline
S\&P 500 & \cellcolor{gray!25} 0.2721 & \cellcolor{gray!25} 0.2727 & \cellcolor{gray!25} 0.2558 & \cellcolor{gray!25} \textcolor{blue}{0.2488} & 0.2692 & 0.2761 & 0.2761 & 0.2810 \\
IXIC & \cellcolor{gray!25} 0.2285 & \cellcolor{gray!25} 0.2263 & \cellcolor{gray!25} 0.2432 & \cellcolor{gray!25} 0.2403 & \cellcolor{gray!25} \textcolor{blue}{0.2172} & \cellcolor{gray!25} 0.2273 & \cellcolor{gray!25} 0.2273 & \cellcolor{gray!25} 0.2207 \\
DJI & \cellcolor{gray!25} 0.3171 & \cellcolor{gray!25} 0.3191 & \cellcolor{gray!25} \textcolor{blue}{0.2785} & \cellcolor{gray!25} 0.2934 & \cellcolor{gray!25} 0.3130 & \cellcolor{gray!25} 0.3212 & \cellcolor{gray!25} 0.3212 & 0.3296 \\
SSEC & \cellcolor{gray!25} 0.1318 & \cellcolor{gray!25} 0.1323 & \cellcolor{gray!25} \textcolor{blue}{0.1262} & \cellcolor{gray!25} 0.1274 & \cellcolor{gray!25} 0.1332 & \cellcolor{gray!25} 0.1341 & \cellcolor{gray!25} 0.1341 & \cellcolor{gray!25} 0.1340 \\
N225 & \cellcolor{gray!25} 0.1328 & \cellcolor{gray!25} 0.1322 & \cellcolor{gray!25} \textcolor{blue}{0.1315} & \cellcolor{gray!25} 0.1369 & \cellcolor{gray!25} 0.1344 & \cellcolor{gray!25} 0.1352 & \cellcolor{gray!25} 0.1352 & \cellcolor{gray!25} 0.1367 \\
HSI & \cellcolor{gray!25} \textcolor{blue}{0.1602} & \cellcolor{gray!25} 0.1606 & \cellcolor{gray!25} 0.1608 & \cellcolor{gray!25} 0.1619 & 0.1626 & \cellcolor{gray!25} 0.1609 & \cellcolor{gray!25} 0.1609 & 0.1646 \\
STOXX50E & \cellcolor{gray!25} \textcolor{blue}{0.2178} & 0.2196 & 0.2307 & 0.2417 & 0.2306 & 0.2329 & 0.2329 & 0.2365 \\
NSEI & \cellcolor{gray!25} 0.2620 & \cellcolor{gray!25} 0.2669 & \cellcolor{gray!25} 0.2618 & \cellcolor{gray!25} 0.2629 & \cellcolor{gray!25} \textcolor{blue}{0.2612} & \cellcolor{gray!25} 0.2671 & \cellcolor{gray!25} 0.2671 & \cellcolor{gray!25} 0.2643 \\
FTSE & 0.2034 & \cellcolor{gray!25} \textcolor{blue}{0.2001} & 0.2246 & 0.2167 & 0.2231 & 0.2240 & 0.2240 & 0.2238 \\
BSESN & \cellcolor{gray!25} 0.2626 & \cellcolor{gray!25} 0.2612 & \cellcolor{gray!25} 0.2631 & 0.2685 & \cellcolor{gray!25} 0.2515 & \cellcolor{gray!25} 0.2590 & \cellcolor{gray!25} 0.2590 & \cellcolor{gray!25} \textcolor{blue}{0.2506} \\
GSPTSE & \cellcolor{gray!25} 0.2658 & \cellcolor{gray!25} \textcolor{blue}{0.2498} & \cellcolor{gray!25} 0.2784 & 0.2866 & \cellcolor{gray!25} 0.2942 & \cellcolor{gray!25} 0.3024 & \cellcolor{gray!25} 0.3024 & \cellcolor{gray!25} 0.3127 \\
FCHI & \cellcolor{gray!25} 0.2304 & \cellcolor{gray!25} 0.2307 & \cellcolor{gray!25} \textcolor{blue}{0.2291} & \cellcolor{gray!25} 0.2341 & \cellcolor{gray!25} 0.2299 & \cellcolor{gray!25} 0.2313 & \cellcolor{gray!25} 0.2313 & \cellcolor{gray!25} 0.2300 \\
RUT & \cellcolor{gray!25} 0.3259 & \cellcolor{gray!25} 0.3284 & \cellcolor{gray!25} 0.3397 & \cellcolor{gray!25} 0.3200 & \cellcolor{gray!25} \textcolor{blue}{0.3191} & \cellcolor{gray!25} 0.3290 & \cellcolor{gray!25} 0.3290 & \cellcolor{gray!25} 0.3304 \\
GDAXI & 0.2431 & \cellcolor{gray!25} 0.2418 & \cellcolor{gray!25} \textcolor{blue}{0.2397} & \cellcolor{gray!25} 0.2425 & 0.2445 & 0.2445 & 0.2445 & 0.2477 \\
KS11 & \cellcolor{gray!25} 0.1602 & \cellcolor{gray!25} 0.1577 & \cellcolor{gray!25} \textcolor{blue}{0.1563} & \cellcolor{gray!25} 0.1569 & \cellcolor{gray!25} 0.1579 & \cellcolor{gray!25} 0.1600 & \cellcolor{gray!25} 0.1600 & \cellcolor{gray!25} 0.1613 \\
SSMI & \cellcolor{gray!25} \textcolor{blue}{0.1644} & \cellcolor{gray!25} 0.1646 & \cellcolor{gray!25} 0.1648 & \cellcolor{gray!25} 0.1673 & \cellcolor{gray!25} 0.1695 & \cellcolor{gray!25} 0.1703 & \cellcolor{gray!25} 0.1703 & \cellcolor{gray!25} 0.1689 \\
AORD & \cellcolor{gray!25} 0.1817 & \cellcolor{gray!25} \textcolor{blue}{0.1811} & \cellcolor{gray!25} 0.1861 & \cellcolor{gray!25} 0.1867 & \cellcolor{gray!25} 0.1853 & \cellcolor{gray!25} 0.1871 & \cellcolor{gray!25} 0.1871 & \cellcolor{gray!25} 0.1868 \\
BVSP & \cellcolor{gray!25} 0.4110 & \cellcolor{gray!25} 0.4098 & \cellcolor{gray!25} \textcolor{blue}{0.3700} & \cellcolor{gray!25} 0.3965 & \cellcolor{gray!25} 0.3832 & \cellcolor{gray!25} 0.4136 & \cellcolor{gray!25} 0.4136 & \cellcolor{gray!25} 0.3877 \\
IBEX & \cellcolor{gray!25} 0.2472 & \cellcolor{gray!25} 0.2465 & \cellcolor{gray!25} \textcolor{blue}{0.2366} & \cellcolor{gray!25} 0.2431 & \cellcolor{gray!25} 0.2431 & \cellcolor{gray!25} 0.2508 & \cellcolor{gray!25} 0.2508 & \cellcolor{gray!25} 0.2471 \\
FTMIB & \cellcolor{gray!25} 0.3284 & \cellcolor{gray!25} 0.3282 & \cellcolor{gray!25} \textcolor{blue}{0.3274} & \cellcolor{gray!25} 0.3314 & \cellcolor{gray!25} 0.3278 & \cellcolor{gray!25} 0.3292 & \cellcolor{gray!25} 0.3292 & \cellcolor{gray!25} 0.3298 \\
AEX & \cellcolor{gray!25} 0.1940 & \cellcolor{gray!25} 0.1945 & \cellcolor{gray!25} \textcolor{blue}{0.1940} & 0.1983 & \cellcolor{gray!25} 0.1962 & \cellcolor{gray!25} 0.1973 & \cellcolor{gray!25} 0.1973 & \cellcolor{gray!25} 0.1975 \\
OMXSPI & \cellcolor{gray!25} 0.2263 & \cellcolor{gray!25} \textcolor{blue}{0.2253} & 0.2308 & \cellcolor{gray!25} 0.2372 & \cellcolor{gray!25} 0.2374 & \cellcolor{gray!25} 0.2382 & \cellcolor{gray!25} 0.2382 & \cellcolor{gray!25} 0.2386 \\
MXX & \cellcolor{gray!25} 0.1777 & \cellcolor{gray!25} 0.1770 & \cellcolor{gray!25} 0.1788 & \cellcolor{gray!25} \textcolor{blue}{0.1737} & \cellcolor{gray!25} 0.1777 & \cellcolor{gray!25} 0.1792 & \cellcolor{gray!25} 0.1792 & 0.1834 \\
BFX & \cellcolor{gray!25} 0.2535 & \cellcolor{gray!25} 0.2536 & \cellcolor{gray!25} \textcolor{blue}{0.2442} & \cellcolor{gray!25} 0.2561 & \cellcolor{gray!25} 0.2485 & \cellcolor{gray!25} 0.2558 & \cellcolor{gray!25} 0.2558 & \cellcolor{gray!25} 0.2535 \\
OSEAX & 0.2198 & \cellcolor{gray!25} 0.2173 & \cellcolor{gray!25} 0.2263 & \cellcolor{gray!25} 0.2223 & \cellcolor{gray!25} \textcolor{blue}{0.2168} & \cellcolor{gray!25} 0.2200 & \cellcolor{gray!25} 0.2200 & \cellcolor{gray!25} 0.2191 \\
OMXC20 & \cellcolor{gray!25} 0.1452 & \cellcolor{gray!25} 0.1455 & \cellcolor{gray!25} \textcolor{blue}{0.1405} & \cellcolor{gray!25} 0.1424 & \cellcolor{gray!25} 0.1475 & \cellcolor{gray!25} 0.1481 & \cellcolor{gray!25} 0.1481 & \cellcolor{gray!25} 0.1505 \\
OMXHPI & \cellcolor{gray!25} 0.2101 & \cellcolor{gray!25} 0.2104 & \cellcolor{gray!25} \textcolor{blue}{0.2094} & \cellcolor{gray!25} 0.2132 & \cellcolor{gray!25} 0.2188 & \cellcolor{gray!25} 0.2181 & \cellcolor{gray!25} 0.2181 & \cellcolor{gray!25} 0.2206 \\
SMSI & \cellcolor{gray!25} \textcolor{blue}{0.2276} & \cellcolor{gray!25} 0.2281 & \cellcolor{gray!25} 0.2311 & \cellcolor{gray!25} 0.2338 & \cellcolor{gray!25} 0.2335 & \cellcolor{gray!25} 0.2363 & \cellcolor{gray!25} 0.2363 & \cellcolor{gray!25} 0.2352 \\
KSE & \cellcolor{gray!25} 0.1715 & \cellcolor{gray!25} 0.1714 & \cellcolor{gray!25} 0.1713 & \cellcolor{gray!25} 0.1711 & \cellcolor{gray!25} 0.1680 & \cellcolor{gray!25} 0.1713 & \cellcolor{gray!25} 0.1713 & \cellcolor{gray!25} \textcolor{blue}{0.1672} \\
\hline
\end{tabular}%
\caption*{\tiny{Note: The table reports average quantile score values. Shaded cells indicate models retained in the MCS. Blue values indicate the lowest average quantile score for each market. Lower values indicate better quantile forecasting performance.}}
\end{table}

%% file: Table/jointloss_combined_table.tex
\clearpage
\begin{table}[htbp]
\centering
\caption{Average AL joint loss and MCS results.}
\label{tab:jointloss_combined}
\tiny
\renewcommand{\arraystretch}{0.8}
\newcolumntype{C}{>{\centering\arraybackslash}p{4.5em}}
\begin{tabular}{lCCCC:CCCC}
\hline
Whole period & \multicolumn{4}{c:}{Proposed Models}
& \multicolumn{4}{c}{Benchmarks} \\
01/01/2018--28/06/2022 & \begin{tabular}{@{}c@{}}AE-RG\\(12-6-1)\end{tabular} & \begin{tabular}{@{}c@{}}AE-RG\\(12-12-1)\end{tabular} & LW-RG & FW-RG & AVG-RG & PC-RG & IC-RG & RG \\
\hline
S\&P 500 & 2.157 & 2.158 & \cellcolor{gray!25} \textcolor{blue}{2.125} & \cellcolor{gray!25} 2.131 & 2.171 & 2.175 & 2.175 & 2.209 \\
IXIC & \cellcolor{gray!25} \textcolor{blue}{2.233} & \cellcolor{gray!25} 2.234 & \cellcolor{gray!25} 2.270 & \cellcolor{gray!25} 2.265 & \cellcolor{gray!25} 2.238 & \cellcolor{gray!25} 2.244 & \cellcolor{gray!25} 2.244 & \cellcolor{gray!25} 2.241 \\
DJI & 2.213 & 2.216 & \cellcolor{gray!25} \textcolor{blue}{2.135} & 2.175 & 2.221 & 2.229 & 2.229 & 2.276 \\
SSEC & \cellcolor{gray!25} 2.262 & \cellcolor{gray!25} 2.225 & \cellcolor{gray!25} \textcolor{blue}{2.188} & 2.205 & \cellcolor{gray!25} 2.220 & 2.282 & 2.282 & \cellcolor{gray!25} 2.233 \\
N225 & 2.257 & 2.254 & \cellcolor{gray!25} \textcolor{blue}{2.237} & 2.272 & 2.256 & 2.265 & 2.265 & 2.270 \\
HSI & \cellcolor{gray!25} 2.361 & \cellcolor{gray!25} \textcolor{blue}{2.359} & \cellcolor{gray!25} 2.366 & 2.382 & \cellcolor{gray!25} 2.370 & \cellcolor{gray!25} 2.363 & \cellcolor{gray!25} 2.363 & \cellcolor{gray!25} 2.378 \\
STOXX50E & \cellcolor{gray!25} \textcolor{blue}{2.273} & 2.281 & 2.335 & 2.354 & 2.336 & 2.349 & 2.349 & 2.349 \\
NSEI & \cellcolor{gray!25} 2.176 & \cellcolor{gray!25} 2.183 & \cellcolor{gray!25} \textcolor{blue}{2.165} & \cellcolor{gray!25} 2.191 & \cellcolor{gray!25} 2.180 & \cellcolor{gray!25} 2.183 & \cellcolor{gray!25} 2.183 & \cellcolor{gray!25} 2.193 \\
FTSE & 2.200 & \cellcolor{gray!25} \textcolor{blue}{2.169} & 2.274 & 2.252 & 2.274 & 2.286 & 2.286 & 2.277 \\
BSESN & \cellcolor{gray!25} \textcolor{blue}{2.132} & 2.139 & \cellcolor{gray!25} 2.133 & \cellcolor{gray!25} 2.138 & \cellcolor{gray!25} 2.137 & \cellcolor{gray!25} 2.134 & \cellcolor{gray!25} 2.134 & \cellcolor{gray!25} 2.148 \\
GSPTSE & \cellcolor{gray!25} 1.911 & \cellcolor{gray!25} 1.917 & \cellcolor{gray!25} \textcolor{blue}{1.871} & 1.906 & 1.933 & 1.959 & 1.959 & 1.990 \\
FCHI & \cellcolor{gray!25} 2.375 & \cellcolor{gray!25} 2.375 & \cellcolor{gray!25} \textcolor{blue}{2.351} & \cellcolor{gray!25} 2.363 & \cellcolor{gray!25} 2.364 & 2.386 & 2.386 & \cellcolor{gray!25} 2.363 \\
RUT & \cellcolor{gray!25} 2.295 & \cellcolor{gray!25} 2.296 & \cellcolor{gray!25} 2.299 & \cellcolor{gray!25} 2.295 & \cellcolor{gray!25} \textcolor{blue}{2.290} & \cellcolor{gray!25} 2.301 & \cellcolor{gray!25} 2.301 & \cellcolor{gray!25} 2.302 \\
GDAXI & 2.471 & 2.468 & 2.479 & \cellcolor{gray!25} \textcolor{blue}{2.449} & 2.473 & 2.479 & 2.479 & 2.489 \\
KS11 & \cellcolor{gray!25} 2.028 & \cellcolor{gray!25} 2.027 & \cellcolor{gray!25} \textcolor{blue}{2.021} & \cellcolor{gray!25} 2.022 & \cellcolor{gray!25} 2.032 & \cellcolor{gray!25} 2.033 & \cellcolor{gray!25} 2.033 & \cellcolor{gray!25} 2.044 \\
SSMI & \cellcolor{gray!25} 2.025 & \cellcolor{gray!25} 2.034 & 2.111 & \cellcolor{gray!25} 2.010 & \cellcolor{gray!25} 2.013 & \cellcolor{gray!25} 2.021 & \cellcolor{gray!25} 2.021 & \cellcolor{gray!25} \textcolor{blue}{2.007} \\
AORD & \cellcolor{gray!25} \textcolor{blue}{2.119} & \cellcolor{gray!25} 2.122 & \cellcolor{gray!25} 2.137 & \cellcolor{gray!25} 2.139 & \cellcolor{gray!25} 2.136 & \cellcolor{gray!25} 2.136 & \cellcolor{gray!25} 2.136 & 2.148 \\
BVSP & \cellcolor{gray!25} 2.473 & \cellcolor{gray!25} 2.470 & \cellcolor{gray!25} \textcolor{blue}{2.446} & \cellcolor{gray!25} 2.469 & \cellcolor{gray!25} 2.460 & \cellcolor{gray!25} 2.474 & \cellcolor{gray!25} 2.474 & \cellcolor{gray!25} 2.464 \\
IBEX & \cellcolor{gray!25} 2.221 & \cellcolor{gray!25} \textcolor{blue}{2.217} & \cellcolor{gray!25} 2.264 & \cellcolor{gray!25} 2.258 & \cellcolor{gray!25} 2.257 & \cellcolor{gray!25} 2.259 & \cellcolor{gray!25} 2.259 & \cellcolor{gray!25} 2.254 \\
FTMIB & \cellcolor{gray!25} 2.487 & \cellcolor{gray!25} 2.488 & \cellcolor{gray!25} 2.516 & \cellcolor{gray!25} 2.470 & \cellcolor{gray!25} \textcolor{blue}{2.469} & \cellcolor{gray!25} 2.481 & \cellcolor{gray!25} 2.481 & \cellcolor{gray!25} 2.476 \\
AEX & \cellcolor{gray!25} 2.190 & \cellcolor{gray!25} 2.190 & 2.210 & 2.220 & 2.201 & 2.203 & 2.203 & \cellcolor{gray!25} \textcolor{blue}{2.189} \\
OMXSPI & \cellcolor{gray!25} \textcolor{blue}{2.167} & \cellcolor{gray!25} 2.171 & \cellcolor{gray!25} 2.211 & \cellcolor{gray!25} 2.233 & \cellcolor{gray!25} 2.229 & \cellcolor{gray!25} 2.228 & \cellcolor{gray!25} 2.228 & 2.249 \\
MXX & 2.070 & \cellcolor{gray!25} 2.068 & 2.081 & \cellcolor{gray!25} 2.064 & \cellcolor{gray!25} \textcolor{blue}{2.058} & \cellcolor{gray!25} 2.059 & \cellcolor{gray!25} 2.059 & \cellcolor{gray!25} 2.065 \\
BFX & \cellcolor{gray!25} 2.133 & \cellcolor{gray!25} \textcolor{blue}{2.132} & \cellcolor{gray!25} 2.134 & 2.152 & \cellcolor{gray!25} 2.134 & 2.133 & \cellcolor{gray!25} 2.133 & \cellcolor{gray!25} 2.146 \\
OSEAX & \cellcolor{gray!25} 2.112 & \cellcolor{gray!25} \textcolor{blue}{2.111} & 2.172 & \cellcolor{gray!25} 2.138 & \cellcolor{gray!25} 2.115 & \cellcolor{gray!25} 2.120 & \cellcolor{gray!25} 2.120 & \cellcolor{gray!25} 2.121 \\
OMXC20 & \cellcolor{gray!25} 2.172 & \cellcolor{gray!25} 2.170 & \cellcolor{gray!25} \textcolor{blue}{2.135} & \cellcolor{gray!25} 2.152 & \cellcolor{gray!25} 2.159 & \cellcolor{gray!25} 2.161 & \cellcolor{gray!25} 2.161 & 2.174 \\
OMXHPI & \cellcolor{gray!25} 2.146 & \cellcolor{gray!25} 2.144 & \cellcolor{gray!25} \textcolor{blue}{2.093} & \cellcolor{gray!25} 2.106 & \cellcolor{gray!25} 2.103 & \cellcolor{gray!25} 2.113 & \cellcolor{gray!25} 2.113 & \cellcolor{gray!25} 2.121 \\
SMSI & \cellcolor{gray!25} 2.192 & \cellcolor{gray!25} \textcolor{blue}{2.192} & \cellcolor{gray!25} 2.215 & \cellcolor{gray!25} 2.213 & \cellcolor{gray!25} 2.220 & \cellcolor{gray!25} 2.243 & \cellcolor{gray!25} 2.243 & \cellcolor{gray!25} 2.222 \\
KSE & \cellcolor{gray!25} 2.235 & \cellcolor{gray!25} 2.236 & \cellcolor{gray!25} 2.238 & \cellcolor{gray!25} 2.238 & \cellcolor{gray!25} 2.227 & \cellcolor{gray!25} 2.236 & \cellcolor{gray!25} 2.236 & \cellcolor{gray!25} \textcolor{blue}{2.226} \\
\hline
COVID period & \multicolumn{4}{c:}{Proposed Models}
& \multicolumn{4}{c}{Benchmarks} \\
30/01/2020--30/06/2020 & \begin{tabular}{@{}c@{}}AE-RG\\(12-6-1)\end{tabular} & \begin{tabular}{@{}c@{}}AE-RG\\(12-12-1)\end{tabular} & LW-RG & FW-RG & AVG-RG & PC-RG & IC-RG & RG \\
\hline
S\&P 500 & 3.927 & 3.939 & 3.743 & \cellcolor{gray!25} \textcolor{blue}{3.670} & 3.948 & 4.004 & 4.004 & 4.088 \\
IXIC & \cellcolor{gray!25} 3.274 & \cellcolor{gray!25} 3.260 & \cellcolor{gray!25} 3.371 & \cellcolor{gray!25} 3.350 & \cellcolor{gray!25} \textcolor{blue}{3.224} & \cellcolor{gray!25} 3.281 & \cellcolor{gray!25} 3.281 & \cellcolor{gray!25} 3.245 \\
DJI & 4.312 & 4.345 & \cellcolor{gray!25} \textcolor{blue}{3.837} & 4.022 & 4.318 & 4.404 & 4.404 & 4.567 \\
SSEC & \cellcolor{gray!25} 2.817 & \cellcolor{gray!25} 2.823 & \cellcolor{gray!25} \textcolor{blue}{2.666} & \cellcolor{gray!25} 2.712 & \cellcolor{gray!25} 2.836 & \cellcolor{gray!25} 2.877 & \cellcolor{gray!25} 2.877 & \cellcolor{gray!25} 2.874 \\
N225 & \cellcolor{gray!25} 2.730 & \cellcolor{gray!25} \textcolor{blue}{2.721} & \cellcolor{gray!25} 2.735 & \cellcolor{gray!25} 2.771 & \cellcolor{gray!25} 2.749 & \cellcolor{gray!25} 2.752 & \cellcolor{gray!25} 2.752 & \cellcolor{gray!25} 2.774 \\
HSI & \cellcolor{gray!25} \textcolor{blue}{3.306} & \cellcolor{gray!25} 3.316 & \cellcolor{gray!25} 3.368 & \cellcolor{gray!25} 3.409 & 3.365 & 3.324 & 3.324 & 3.421 \\
STOXX50E & \cellcolor{gray!25} \textcolor{blue}{3.306} & 3.326 & 3.498 & 3.664 & 3.504 & 3.538 & 3.538 & 3.571 \\
NSEI & \cellcolor{gray!25} 3.990 & \cellcolor{gray!25} 4.013 & \cellcolor{gray!25} \textcolor{blue}{3.976} & 4.040 & \cellcolor{gray!25} 4.013 & \cellcolor{gray!25} 4.033 & \cellcolor{gray!25} 4.033 & \cellcolor{gray!25} 4.058 \\
FTSE & 3.280 & \cellcolor{gray!25} \textcolor{blue}{3.242} & 3.673 & 3.502 & 3.599 & 3.600 & 3.600 & 3.590 \\
BSESN & \cellcolor{gray!25} 3.886 & \cellcolor{gray!25} 3.908 & \cellcolor{gray!25} 3.906 & \cellcolor{gray!25} 3.942 & \cellcolor{gray!25} 3.867 & \cellcolor{gray!25} 3.888 & \cellcolor{gray!25} 3.888 & \cellcolor{gray!25} \textcolor{blue}{3.864} \\
GSPTSE & 3.475 & \cellcolor{gray!25} \textcolor{blue}{3.364} & 4.192 & 4.369 & 4.647 & 4.848 & 4.848 & 4.992 \\
FCHI & \cellcolor{gray!25} 3.455 & \cellcolor{gray!25} 3.462 & \cellcolor{gray!25} \textcolor{blue}{3.422} & \cellcolor{gray!25} 3.500 & \cellcolor{gray!25} 3.438 & \cellcolor{gray!25} 3.470 & \cellcolor{gray!25} 3.470 & \cellcolor{gray!25} 3.427 \\
RUT & \cellcolor{gray!25} 3.756 & \cellcolor{gray!25} 3.779 & \cellcolor{gray!25} 3.875 & \cellcolor{gray!25} \textcolor{blue}{3.705} & \cellcolor{gray!25} 3.724 & \cellcolor{gray!25} 3.822 & \cellcolor{gray!25} 3.822 & \cellcolor{gray!25} 3.815 \\
GDAXI & \cellcolor{gray!25} 3.717 & \cellcolor{gray!25} 3.709 & \cellcolor{gray!25} 3.686 & \cellcolor{gray!25} \textcolor{blue}{3.674} & 3.737 & 3.750 & 3.750 & 3.799 \\
KS11 & \cellcolor{gray!25} 2.839 & \cellcolor{gray!25} 2.817 & \cellcolor{gray!25} \textcolor{blue}{2.804} & \cellcolor{gray!25} 2.809 & \cellcolor{gray!25} 2.829 & \cellcolor{gray!25} 2.843 & \cellcolor{gray!25} 2.843 & \cellcolor{gray!25} 2.857 \\
SSMI & \cellcolor{gray!25} \textcolor{blue}{3.007} & \cellcolor{gray!25} 3.038 & \cellcolor{gray!25} 3.055 & \cellcolor{gray!25} 3.040 & \cellcolor{gray!25} 3.073 & \cellcolor{gray!25} 3.086 & \cellcolor{gray!25} 3.086 & \cellcolor{gray!25} 3.046 \\
AORD & \cellcolor{gray!25} 3.296 & \cellcolor{gray!25} \textcolor{blue}{3.294} & \cellcolor{gray!25} 3.372 & \cellcolor{gray!25} 3.378 & \cellcolor{gray!25} 3.375 & \cellcolor{gray!25} 3.412 & \cellcolor{gray!25} 3.412 & \cellcolor{gray!25} 3.398 \\
BVSP & \cellcolor{gray!25} 4.364 & \cellcolor{gray!25} 4.334 & \cellcolor{gray!25} \textcolor{blue}{4.076} & \cellcolor{gray!25} 4.304 & \cellcolor{gray!25} 4.207 & \cellcolor{gray!25} 4.397 & \cellcolor{gray!25} 4.397 & \cellcolor{gray!25} 4.244 \\
IBEX & \cellcolor{gray!25} 3.470 & \cellcolor{gray!25} \textcolor{blue}{3.457} & \cellcolor{gray!25} 3.511 & \cellcolor{gray!25} 3.557 & \cellcolor{gray!25} 3.611 & \cellcolor{gray!25} 3.673 & \cellcolor{gray!25} 3.673 & \cellcolor{gray!25} 3.660 \\
FTMIB & \cellcolor{gray!25} 3.936 & \cellcolor{gray!25} 3.936 & \cellcolor{gray!25} 3.971 & \cellcolor{gray!25} 3.991 & \cellcolor{gray!25} \textcolor{blue}{3.922} & \cellcolor{gray!25} 3.949 & \cellcolor{gray!25} 3.949 & \cellcolor{gray!25} 3.961 \\
AEX & \cellcolor{gray!25} \textcolor{blue}{3.192} & \cellcolor{gray!25} 3.198 & \cellcolor{gray!25} 3.213 & \cellcolor{gray!25} 3.264 & \cellcolor{gray!25} 3.225 & \cellcolor{gray!25} 3.244 & \cellcolor{gray!25} 3.244 & \cellcolor{gray!25} 3.244 \\
OMXSPI & \cellcolor{gray!25} 3.610 & \cellcolor{gray!25} \textcolor{blue}{3.600} & 3.685 & 3.770 & 3.761 & 3.773 & 3.773 & 3.770 \\
MXX & \cellcolor{gray!25} 3.123 & \cellcolor{gray!25} 3.108 & \cellcolor{gray!25} 3.132 & \cellcolor{gray!25} \textcolor{blue}{3.074} & \cellcolor{gray!25} 3.133 & \cellcolor{gray!25} 3.159 & \cellcolor{gray!25} 3.159 & 3.212 \\
BFX & \cellcolor{gray!25} 3.562 & \cellcolor{gray!25} 3.568 & \cellcolor{gray!25} \textcolor{blue}{3.483} & \cellcolor{gray!25} 3.592 & \cellcolor{gray!25} 3.505 & \cellcolor{gray!25} 3.593 & \cellcolor{gray!25} 3.593 & \cellcolor{gray!25} 3.574 \\
OSEAX & 3.470 & \cellcolor{gray!25} 3.433 & \cellcolor{gray!25} 3.673 & \cellcolor{gray!25} 3.575 & \cellcolor{gray!25} 3.413 & \cellcolor{gray!25} 3.440 & \cellcolor{gray!25} 3.440 & \cellcolor{gray!25} \textcolor{blue}{3.394} \\
OMXC20 & \cellcolor{gray!25} 2.803 & \cellcolor{gray!25} 2.799 & \cellcolor{gray!25} 2.756 & \cellcolor{gray!25} \textcolor{blue}{2.753} & \cellcolor{gray!25} 2.832 & \cellcolor{gray!25} 2.841 & \cellcolor{gray!25} 2.841 & \cellcolor{gray!25} 2.876 \\
OMXHPI & \cellcolor{gray!25} \textcolor{blue}{3.348} & \cellcolor{gray!25} 3.349 & \cellcolor{gray!25} 3.390 & \cellcolor{gray!25} 3.435 & \cellcolor{gray!25} 3.526 & \cellcolor{gray!25} 3.518 & \cellcolor{gray!25} 3.518 & \cellcolor{gray!25} 3.536 \\
SMSI & \cellcolor{gray!25} 3.304 & \cellcolor{gray!25} \textcolor{blue}{3.301} & \cellcolor{gray!25} 3.420 & \cellcolor{gray!25} 3.464 & \cellcolor{gray!25} 3.458 & \cellcolor{gray!25} 3.512 & \cellcolor{gray!25} 3.512 & \cellcolor{gray!25} 3.495 \\
KSE & \cellcolor{gray!25} 3.216 & \cellcolor{gray!25} 3.216 & \cellcolor{gray!25} 3.220 & \cellcolor{gray!25} 3.215 & \cellcolor{gray!25} 3.159 & \cellcolor{gray!25} 3.218 & \cellcolor{gray!25} 3.218 & \cellcolor{gray!25} \textcolor{blue}{3.140} \\
\hline
\end{tabular}%
\caption*{\tiny{Note: The table reports average AL joint loss values. Shaded cells indicate models retained in the MCS. Blue values indicate the lowest average AL joint loss for each market. Lower values indicate better joint VaR and ES forecasting performance.}}

\end{table}

%% file: Table/pnll_combined_table_1dp.tex
\clearpage
\begin{table}[H]
\centering
\caption{Negative predictive log-likelihood and MCS results.}
\label{tab:pnll_combined_1dp}
\tiny
\setlength{\tabcolsep}{5pt}
\renewcommand{\arraystretch}{0.8}
\newcolumntype{C}{>{\centering\arraybackslash}p{4.8em}}
\begin{tabular}{lCCCC:CCCC}
\hline
Whole period & \multicolumn{4}{c:}{Proposed Models}
& \multicolumn{4}{c}{Benchmarks} \\
01/01/2018--28/06/2022 & \begin{tabular}{@{}c@{}}AE-RG\\(12-6-1)\end{tabular} & \begin{tabular}{@{}c@{}}AE-RG\\(12-12-1)\end{tabular} & LW-RG & FW-RG & AVG-RG & PC-RG & IC-RG & RG \\
\hline
S\&P 500 & 992.0 & 993.1 & \cellcolor{gray!25} \textcolor{blue}{970.4} & \cellcolor{gray!25} 974.7 & 992.6 & 1000.3 & 1000.3 & 1011.1 \\
IXIC & 1542.0 & 1540.8 & \cellcolor{gray!25} 1515.9 & \cellcolor{gray!25} 1514.7 & \cellcolor{gray!25} \textcolor{blue}{1506.9} & \cellcolor{gray!25} 1525.1 & \cellcolor{gray!25} 1525.1 & 1512.2 \\
DJI & 1030.3 & 1032.1 & \cellcolor{gray!25} \textcolor{blue}{990.3} & 1009.3 & 1032.4 & 1040.3 & 1040.3 & 1065.5 \\
SSEC & \cellcolor{gray!25} 1270.3 & \cellcolor{gray!25} 1250.7 & 1301.1 & 1315.5 & \cellcolor{gray!25} \textcolor{blue}{1246.5} & \cellcolor{gray!25} 1273.2 & \cellcolor{gray!25} 1273.2 & \cellcolor{gray!25} 1252.7 \\
N225 & 1406.8 & 1406.0 & \cellcolor{gray!25} \textcolor{blue}{1380.8} & 1406.3 & 1393.7 & 1407.5 & 1407.5 & 1405.0 \\
HSI & \cellcolor{gray!25} 1627.7 & \cellcolor{gray!25} 1627.8 & \cellcolor{gray!25} \textcolor{blue}{1620.2} & \cellcolor{gray!25} 1624.0 & \cellcolor{gray!25} 1624.6 & \cellcolor{gray!25} 1626.2 & \cellcolor{gray!25} 1626.2 & \cellcolor{gray!25} 1625.1 \\
STOXX50E & \cellcolor{gray!25} \textcolor{blue}{1264.9} & \cellcolor{gray!25} 1269.2 & \cellcolor{gray!25} 1275.7 & \cellcolor{gray!25} 1281.7 & \cellcolor{gray!25} 1272.6 & \cellcolor{gray!25} 1284.0 & \cellcolor{gray!25} 1284.0 & 1281.8 \\
NSEI & \cellcolor{gray!25} 1094.4 & \cellcolor{gray!25} 1095.9 & \cellcolor{gray!25} \textcolor{blue}{1090.7} & \cellcolor{gray!25} 1097.1 & \cellcolor{gray!25} 1091.3 & \cellcolor{gray!25} 1093.7 & \cellcolor{gray!25} 1093.7 & \cellcolor{gray!25} 1095.7 \\
FTSE & \cellcolor{gray!25} 1035.1 & \cellcolor{gray!25} \textcolor{blue}{1023.9} & \cellcolor{gray!25} 1047.1 & \cellcolor{gray!25} 1037.1 & \cellcolor{gray!25} 1040.9 & \cellcolor{gray!25} 1047.3 & \cellcolor{gray!25} 1047.3 & \cellcolor{gray!25} 1046.3 \\
BSESN & \cellcolor{gray!25} \textcolor{blue}{1071.1} & \cellcolor{gray!25} 1072.4 & \cellcolor{gray!25} 1072.9 & \cellcolor{gray!25} 1074.5 & \cellcolor{gray!25} 1072.2 & \cellcolor{gray!25} 1073.9 & \cellcolor{gray!25} 1073.9 & \cellcolor{gray!25} 1079.6 \\
GSPTSE & 636.3 & 641.2 & \cellcolor{gray!25} \textcolor{blue}{287.3} & \cellcolor{gray!25} 296.3 & \cellcolor{gray!25} 303.9 & \cellcolor{gray!25} 315.2 & \cellcolor{gray!25} 315.2 & 334.1 \\
FCHI & \cellcolor{gray!25} 1254.2 & \cellcolor{gray!25} 1255.0 & \cellcolor{gray!25} \textcolor{blue}{1243.9} & 1253.6 & \cellcolor{gray!25} 1249.0 & 1258.7 & 1258.7 & \cellcolor{gray!25} 1251.7 \\
RUT & 1729.0 & \cellcolor{gray!25} 1728.5 & \cellcolor{gray!25} 1722.4 & 1731.1 & \cellcolor{gray!25} \textcolor{blue}{1721.4} & \cellcolor{gray!25} 1727.2 & \cellcolor{gray!25} 1727.2 & 1735.5 \\
GDAXI & \cellcolor{gray!25} 1402.2 & \cellcolor{gray!25} 1401.9 & \cellcolor{gray!25} 1404.0 & \cellcolor{gray!25} \textcolor{blue}{1395.8} & \cellcolor{gray!25} 1404.9 & 1407.1 & 1407.1 & 1413.6 \\
KS11 & \cellcolor{gray!25} 1023.3 & \cellcolor{gray!25} 1022.5 & \cellcolor{gray!25} 1023.0 & \cellcolor{gray!25} 1023.2 & \cellcolor{gray!25} \textcolor{blue}{1020.9} & \cellcolor{gray!25} 1024.5 & \cellcolor{gray!25} 1024.5 & \cellcolor{gray!25} 1025.7 \\
SSMI & \cellcolor{gray!25} 755.9 & \cellcolor{gray!25} 764.7 & 788.0 & \cellcolor{gray!25} 753.6 & \cellcolor{gray!25} \textcolor{blue}{751.2} & \cellcolor{gray!25} 759.1 & \cellcolor{gray!25} 759.1 & \cellcolor{gray!25} 751.6 \\
AORD & \cellcolor{gray!25} 719.7 & \cellcolor{gray!25} 720.0 & \cellcolor{gray!25} 725.1 & \cellcolor{gray!25} 724.5 & \cellcolor{gray!25} \textcolor{blue}{719.3} & \cellcolor{gray!25} 722.1 & \cellcolor{gray!25} 722.1 & 726.7 \\
BVSP & \cellcolor{gray!25} 1868.0 & \cellcolor{gray!25} 1877.9 & \cellcolor{gray!25} 1844.4 & \cellcolor{gray!25} 1848.0 & \cellcolor{gray!25} \textcolor{blue}{1842.8} & \cellcolor{gray!25} 1871.9 & \cellcolor{gray!25} 1871.9 & \cellcolor{gray!25} 1845.2 \\
IBEX & 1328.7 & 1327.3 & \cellcolor{gray!25} 1290.8 & \cellcolor{gray!25} 1290.4 & \cellcolor{gray!25} \textcolor{blue}{1288.3} & 1292.8 & 1292.8 & \cellcolor{gray!25} 1288.4 \\
FTMIB & \cellcolor{gray!25} 1538.6 & \cellcolor{gray!25} 1539.1 & \cellcolor{gray!25} 1536.8 & \cellcolor{gray!25} 1528.8 & \cellcolor{gray!25} \textcolor{blue}{1522.1} & \cellcolor{gray!25} 1534.2 & \cellcolor{gray!25} 1534.2 & \cellcolor{gray!25} 1529.9 \\
AEX & \cellcolor{gray!25} 1080.3 & \cellcolor{gray!25} \textcolor{blue}{1078.9} & \cellcolor{gray!25} 1087.1 & 1091.9 & \cellcolor{gray!25} 1080.7 & 1086.9 & 1086.9 & \cellcolor{gray!25} 1080.8 \\
OMXSPI & \cellcolor{gray!25} 1181.2 & \cellcolor{gray!25} 1180.7 & \cellcolor{gray!25} \textcolor{blue}{1156.5} & \cellcolor{gray!25} 1163.5 & \cellcolor{gray!25} 1159.2 & \cellcolor{gray!25} 1161.8 & \cellcolor{gray!25} 1161.8 & 1171.0 \\
MXX & 1208.3 & 1207.2 & 1221.6 & 1205.1 & \cellcolor{gray!25} 1192.6 & \cellcolor{gray!25} 1191.5 & \cellcolor{gray!25} \textcolor{blue}{1191.5} & 1198.5 \\
BFX & 1140.3 & 1141.5 & \cellcolor{gray!25} \textcolor{blue}{1128.9} & 1143.7 & \cellcolor{gray!25} 1133.3 & 1142.1 & 1142.1 & 1147.3 \\
OSEAX & \cellcolor{gray!25} \textcolor{blue}{1143.6} & \cellcolor{gray!25} 1144.0 & \cellcolor{gray!25} 1153.0 & \cellcolor{gray!25} 1149.4 & \cellcolor{gray!25} 1147.5 & 1155.2 & 1155.2 & 1161.5 \\
OMXC20 & 1411.5 & 1412.3 & \cellcolor{gray!25} \textcolor{blue}{1320.6} & 1335.5 & 1331.5 & 1355.0 & 1355.0 & 1341.7 \\
OMXHPI & 1250.4 & 1252.8 & \cellcolor{gray!25} \textcolor{blue}{1123.0} & \cellcolor{gray!25} 1127.2 & \cellcolor{gray!25} 1125.0 & 1132.2 & 1132.2 & 1138.6 \\
SMSI & 1341.1 & 1339.3 & \cellcolor{gray!25} 1276.2 & \cellcolor{gray!25} 1277.6 & \cellcolor{gray!25} 1274.0 & \cellcolor{gray!25} 1284.1 & \cellcolor{gray!25} 1284.1 & \cellcolor{gray!25} \textcolor{blue}{1272.1} \\
KSE & 1317.9 & 1318.6 & 1306.2 & 1304.0 & 1308.3 & 1318.7 & 1318.7 & \cellcolor{gray!25} \textcolor{blue}{1294.7} \\
\hline
COVID period & \multicolumn{4}{c:}{Proposed Models}
& \multicolumn{4}{c}{Benchmarks} \\
30/01/2020--30/06/2020 & \begin{tabular}{@{}c@{}}AE-RG\\(12-6-1)\end{tabular} & \begin{tabular}{@{}c@{}}AE-RG\\(12-12-1)\end{tabular} & LW-RG & FW-RG & AVG-RG & PC-RG & IC-RG & RG \\
\hline
S\&P 500 & 305.2 & 305.7 & 297.7 & \cellcolor{gray!25} \textcolor{blue}{293.0} & 305.8 & 309.0 & 309.0 & 312.1 \\
IXIC & \cellcolor{gray!25} 285.2 & \cellcolor{gray!25} 284.4 & \cellcolor{gray!25} 288.3 & \cellcolor{gray!25} 287.5 & \cellcolor{gray!25} \textcolor{blue}{283.2} & \cellcolor{gray!25} 285.0 & \cellcolor{gray!25} 285.0 & \cellcolor{gray!25} 283.8 \\
DJI & 349.7 & 350.4 & \cellcolor{gray!25} \textcolor{blue}{327.5} & \cellcolor{gray!25} 330.4 & 349.0 & 353.9 & 353.9 & 361.3 \\
SSEC & \cellcolor{gray!25} 144.5 & \cellcolor{gray!25} 144.3 & \cellcolor{gray!25} \textcolor{blue}{142.7} & \cellcolor{gray!25} 146.7 & \cellcolor{gray!25} 144.2 & \cellcolor{gray!25} 146.7 & \cellcolor{gray!25} 146.7 & \cellcolor{gray!25} 147.6 \\
N225 & 244.4 & \cellcolor{gray!25} \textcolor{blue}{244.1} & \cellcolor{gray!25} 244.3 & 245.6 & 245.1 & 245.2 & 245.2 & 246.0 \\
HSI & \cellcolor{gray!25} \textcolor{blue}{225.4} & \cellcolor{gray!25} 225.9 & \cellcolor{gray!25} 226.6 & \cellcolor{gray!25} 228.9 & \cellcolor{gray!25} 226.9 & \cellcolor{gray!25} 226.0 & \cellcolor{gray!25} 226.0 & 230.0 \\
STOXX50E & \cellcolor{gray!25} \textcolor{blue}{296.5} & 298.0 & 305.8 & 312.1 & 308.1 & 309.3 & 309.3 & 310.4 \\
NSEI & \cellcolor{gray!25} 263.5 & \cellcolor{gray!25} 264.2 & \cellcolor{gray!25} \textcolor{blue}{263.5} & 265.9 & \cellcolor{gray!25} 264.2 & \cellcolor{gray!25} 264.9 & \cellcolor{gray!25} 264.9 & \cellcolor{gray!25} 266.2 \\
FTSE & 276.0 & \cellcolor{gray!25} \textcolor{blue}{274.9} & 294.4 & 286.5 & 289.0 & 287.8 & 287.8 & 286.6 \\
BSESN & \cellcolor{gray!25} \textcolor{blue}{275.4} & 277.2 & \cellcolor{gray!25} 277.5 & 278.6 & \cellcolor{gray!25} 275.8 & \cellcolor{gray!25} 276.0 & \cellcolor{gray!25} 276.0 & \cellcolor{gray!25} 275.9 \\
GSPTSE & \cellcolor{gray!25} 253.5 & \cellcolor{gray!25} \textcolor{blue}{247.2} & \cellcolor{gray!25} 278.2 & 287.8 & \cellcolor{gray!25} 305.5 & \cellcolor{gray!25} 317.4 & \cellcolor{gray!25} 317.4 & 325.2 \\
FCHI & \cellcolor{gray!25} 299.9 & \cellcolor{gray!25} 300.3 & \cellcolor{gray!25} \textcolor{blue}{298.2} & \cellcolor{gray!25} 301.5 & \cellcolor{gray!25} 299.6 & \cellcolor{gray!25} 300.7 & \cellcolor{gray!25} 300.7 & \cellcolor{gray!25} 298.2 \\
RUT & 361.1 & 361.3 & 366.6 & \cellcolor{gray!25} \textcolor{blue}{356.7} & \cellcolor{gray!25} 357.6 & 361.9 & 361.9 & 362.3 \\
GDAXI & 325.9 & 326.1 & \cellcolor{gray!25} 324.2 & \cellcolor{gray!25} \textcolor{blue}{322.1} & 327.7 & 328.5 & 328.5 & 331.2 \\
KS11 & \cellcolor{gray!25} 235.9 & \cellcolor{gray!25} 235.1 & \cellcolor{gray!25} 235.1 & \cellcolor{gray!25} \textcolor{blue}{235.0} & \cellcolor{gray!25} 235.7 & \cellcolor{gray!25} 235.8 & \cellcolor{gray!25} 235.8 & \cellcolor{gray!25} 236.2 \\
SSMI & \cellcolor{gray!25} \textcolor{blue}{226.5} & 229.2 & 237.8 & 233.6 & 233.8 & 233.7 & 233.7 & 232.7 \\
AORD & \cellcolor{gray!25} 248.6 & \cellcolor{gray!25} \textcolor{blue}{248.4} & \cellcolor{gray!25} 252.9 & \cellcolor{gray!25} 251.7 & \cellcolor{gray!25} 250.9 & \cellcolor{gray!25} 252.0 & \cellcolor{gray!25} 252.0 & \cellcolor{gray!25} 251.4 \\
BVSP & \cellcolor{gray!25} 364.0 & \cellcolor{gray!25} 362.2 & \cellcolor{gray!25} \textcolor{blue}{349.0} & \cellcolor{gray!25} 362.1 & \cellcolor{gray!25} 356.7 & \cellcolor{gray!25} 365.8 & \cellcolor{gray!25} 365.8 & \cellcolor{gray!25} 359.2 \\
IBEX & 293.2 & \cellcolor{gray!25} \textcolor{blue}{292.1} & \cellcolor{gray!25} 296.4 & \cellcolor{gray!25} 297.4 & \cellcolor{gray!25} 300.1 & \cellcolor{gray!25} 301.8 & \cellcolor{gray!25} 301.8 & \cellcolor{gray!25} 301.6 \\
FTMIB & \cellcolor{gray!25} 317.4 & \cellcolor{gray!25} 317.5 & \cellcolor{gray!25} 317.2 & \cellcolor{gray!25} 321.2 & \cellcolor{gray!25} \textcolor{blue}{316.9} & \cellcolor{gray!25} 318.5 & \cellcolor{gray!25} 318.5 & \cellcolor{gray!25} 319.6 \\
AEX & \cellcolor{gray!25} \textcolor{blue}{276.6} & \cellcolor{gray!25} 276.6 & \cellcolor{gray!25} 278.5 & \cellcolor{gray!25} 279.9 & 278.9 & \cellcolor{gray!25} 279.1 & \cellcolor{gray!25} 279.1 & \cellcolor{gray!25} 279.2 \\
OMXSPI & 276.6 & \cellcolor{gray!25} \textcolor{blue}{275.5} & 279.7 & 282.5 & 282.0 & 282.8 & 282.8 & 282.1 \\
MXX & 234.9 & \cellcolor{gray!25} 234.1 & \cellcolor{gray!25} 235.4 & \cellcolor{gray!25} \textcolor{blue}{232.2} & \cellcolor{gray!25} 235.6 & \cellcolor{gray!25} 236.9 & \cellcolor{gray!25} 236.9 & 239.1 \\
BFX & \cellcolor{gray!25} 318.4 & \cellcolor{gray!25} 318.8 & \cellcolor{gray!25} \textcolor{blue}{318.0} & \cellcolor{gray!25} 320.9 & \cellcolor{gray!25} 319.4 & 320.3 & 320.3 & \cellcolor{gray!25} 320.8 \\
OSEAX & 257.8 & \cellcolor{gray!25} \textcolor{blue}{256.4} & \cellcolor{gray!25} 269.3 & \cellcolor{gray!25} 263.6 & \cellcolor{gray!25} 258.1 & \cellcolor{gray!25} 258.7 & \cellcolor{gray!25} 258.7 & \cellcolor{gray!25} 258.8 \\
OMXC20 & \cellcolor{gray!25} 193.9 & \cellcolor{gray!25} \textcolor{blue}{193.6} & \cellcolor{gray!25} 194.7 & \cellcolor{gray!25} 195.0 & \cellcolor{gray!25} 193.9 & \cellcolor{gray!25} 193.9 & \cellcolor{gray!25} 193.9 & \cellcolor{gray!25} 195.0 \\
OMXHPI & \cellcolor{gray!25} \textcolor{blue}{267.1} & \cellcolor{gray!25} 267.3 & \cellcolor{gray!25} 270.0 & \cellcolor{gray!25} 273.1 & \cellcolor{gray!25} 275.8 & \cellcolor{gray!25} 276.1 & \cellcolor{gray!25} 276.1 & \cellcolor{gray!25} 277.1 \\
SMSI & \cellcolor{gray!25} 286.0 & \cellcolor{gray!25} \textcolor{blue}{285.4} & \cellcolor{gray!25} 293.5 & \cellcolor{gray!25} 294.8 & \cellcolor{gray!25} 295.2 & \cellcolor{gray!25} 297.6 & \cellcolor{gray!25} 297.6 & \cellcolor{gray!25} 297.3 \\
KSE & \cellcolor{gray!25} 207.4 & \cellcolor{gray!25} 207.7 & \cellcolor{gray!25} 205.9 & \cellcolor{gray!25} 205.7 & \cellcolor{gray!25} 204.1 & \cellcolor{gray!25} 208.0 & \cellcolor{gray!25} 208.0 & \cellcolor{gray!25} \textcolor{blue}{202.2} \\
\hline
\end{tabular}%
\caption*{\tiny{Note: The table reports NPLL values. Shaded cells indicate models retained in the MCS. Blue values indicate the lowest NPLL for each market. Lower values indicate better volatility forecasting performance.}}
\end{table}

%% file: references.bib
@article{hansen2011,
	title = {The model confidence set},
	volume = {79},
	issn = {0012-9682},
	url = {https://www.jstor.org/stable/41057463},
	pages = {453--497},
	number = {2},
	journaltitle = {Econometrica},
	publisher = {[Wiley, Econometric Society]},
	author = {Hansen, Peter R. and Lunde, Asger and Nason, James M.},
	urldate = {2026-07-27},
	date = {2011},
	note = {{JCR分区}: Q1
影响因子: 6.2
{SSCI}: Q1
{AJG}: 4*
{ABDC}: A*},
}

@misc{kingma2017,
	title = {Adam: A method for stochastic optimization},
	url = {http://arxiv.org/abs/1412.6980},
	doi = {10.48550/arXiv.1412.6980},
	shorttitle = {Adam},
	number = {{arXiv}:1412.6980},
	publisher = {{arXiv}},
	author = {Kingma, Diederik P. and Ba, Jimmy},
	urldate = {2026-07-06},
	date = {2017-01-30},
	eprinttype = {arxiv},
	eprint = {1412.6980 [cs.LG]},
}

@online{who2026,
	title = {Archived: {WHO} timeline - {COVID}-19},
	url = {https://www.who.int/news/item/27-04-2020-who-timeline---covid-19},
	shorttitle = {Archived},
	author = {{World Health Organization}},
	urldate = {2026-07-03},
	date = {2026-07},
	langid = {english},
}

@article{naimoli2022,
	title = {Improving the accuracy of tail risk forecasting models by combining several realized volatility estimators},
	volume = {107},
	issn = {02649993},
	url = {https://linkinghub.elsevier.com/retrieve/pii/S026499932100290X},
	doi = {10.1016/j.econmod.2021.105701},
	pages = {105701},
	journaltitle = {Economic Modelling},
	shortjournal = {Economic Modelling},
	author = {Naimoli, Antonio and Gerlach, Richard and Storti, Giuseppe},
	urldate = {2024-07-02},
	date = {2022-02},
	langid = {english},
	note = {{JCR分区}: Q1
影响因子: 4.7
{SSCI}: Q1
{ABDC}: A
5年影响因子: 4.8
中科院升级版小类分区: 经济学2区。
{AJG}: 2},
}

@article{hauzenberger2023,
	title = {Real-time inflation forecasting using non-linear dimension reduction techniques},
	volume = {39},
	issn = {01692070},
	url = {https://linkinghub.elsevier.com/retrieve/pii/S0169207022000425},
	doi = {10.1016/j.ijforecast.2022.03.002},
	pages = {901--921},
	number = {2},
	journaltitle = {International Journal of Forecasting},
	shortjournal = {International Journal of Forecasting},
	author = {Hauzenberger, Niko and Huber, Florian and Klieber, Karin},
	urldate = {2026-06-05},
	date = {2023-04},
	note = {{JCR分区}: Q1
影响因子: 5.7
5年影响因子: 7.5
中科院升级版小类分区: 经济学2区/管理学3区。
{SSCI}: Q1
{AJG}: 3
{ABDC}: A},
}

@report{gerdheber2009,
	title = {Oxford-man institute's realized library},
	number = {Version 0.3},
	institution = {Oxford-Man Institute, University of Oxford},
	type = {Data set},
	author = {Heber, Gerd and Lunde, Asger and Shephard, Neil and Sheppard, Kevin K.},
	date = {2009},
}

@article{patton2009,
	title = {Optimal combinations of realised volatility estimators},
	volume = {25},
	issn = {01692070},
	url = {https://linkinghub.elsevier.com/retrieve/pii/S0169207009000107},
	doi = {10.1016/j.ijforecast.2009.01.011},
	series = {Forecasting Returns and Risk in Financial Markets using Linear and Nonlinear Models},
	pages = {218--238},
	number = {2},
	journaltitle = {International Journal of Forecasting},
	shortjournal = {Int. J. Forecasting},
	author = {Patton, Andrew J. and Sheppard, Kevin},
	urldate = {2026-04-29},
	date = {2009-04},
	langid = {american},
	note = {{JCR分区}: Q1
影响因子: 7.1
5年影响因子: 7.5
中科院升级版小类分区: 经济学2区/管理学3区。
{SSCI}: Q1
{AJG}: 3
{ABDC}: A},
}

@article{gu2021,
	title = {Autoencoder asset pricing models},
	volume = {222},
	issn = {03044076},
	url = {https://linkinghub.elsevier.com/retrieve/pii/S0304407620301998},
	doi = {10.1016/j.jeconom.2020.07.009},
	pages = {429--450},
	number = {1},
	journaltitle = {Journal of Econometrics},
	shortjournal = {J. Econometrics},
	author = {Gu, Shihao and Kelly, Bryan and Xiu, Dacheng},
	urldate = {2024-07-02},
	date = {2021-05},
	langid = {english},
	note = {{JCR分区}: Q1
影响因子: 4.0
{SSCI}: Q1
{ABDC}: A*
5年影响因子: 7.8
中科院升级版小类分区: 数学跨学科应用1区/社会科学：数理方法1区/经济学2区。
{AJG}: 4},
}

@article{storti2022,
	title = {Nonparametric expected shortfall forecasting incorporating weighted quantiles},
	volume = {38},
	issn = {01692070},
	url = {https://linkinghub.elsevier.com/retrieve/pii/S0169207021000674},
	doi = {10.1016/j.ijforecast.2021.04.004},
	pages = {224--239},
	number = {1},
	journaltitle = {International Journal of Forecasting},
	shortjournal = {Int. J. Forecasting},
	author = {Storti, Giuseppe and Wang, Chao},
	urldate = {2026-03-03},
	date = {2022-01},
	langid = {english},
	note = {{JCR分区}: Q1
影响因子: 7.1
{SSCI}: Q1
{ABDC}: A
5年影响因子: 7.5
中科院升级版小类分区: 经济学2区/管理学3区。
{AJG}: 3},
}

@report{baker2019,
	title = {Policy News and Stock Market Volatility},
	url = {https://econpapers.repec.org/paper/nbrnberwo/25720.htm},
	number = {25720},
	institution = {National Bureau of Economic Research, Inc},
	type = {{NBER} Working Paper},
	author = {Baker, Scott R and Bloom, Nicholas and Davis, Steven and Kost, Kyle J.},
	urldate = {2024-10-21},
	date = {2019-03},
	langid = {american},
}

@article{taylor2019,
	title = {Forecasting value at risk and expected shortfall using a semiparametric approach based on the asymmetric laplace distribution},
	volume = {37},
	issn = {0735-0015, 1537-2707},
	url = {https://www.tandfonline.com/doi/full/10.1080/07350015.2017.1281815},
	doi = {10.1080/07350015.2017.1281815},
	pages = {121--133},
	number = {1},
	journaltitle = {Journal of Business \& Economic Statistics},
	shortjournal = {Journal of Business \& Economic Statistics},
	publisher = {Taylor \& Francis},
	author = {Taylor, James W.},
	urldate = {2026-03-25},
	date = {2019-01-02},
	note = {\_eprint: https://doi.org/10.1080/07350015.2017.1281815
{JCR分区}: Q1
影响因子: 2.5
5年影响因子: 5.1
中科院升级版小类分区: 统计学与概率论2区/经济学3区/社会科学：数理方法3区。
{SSCI}: Q1
{AJG}: 4
{ABDC}: A*},
}

@article{taylor2020a,
	title = {Forecast combinations for value at risk and expected shortfall},
	volume = {36},
	issn = {01692070},
	url = {https://linkinghub.elsevier.com/retrieve/pii/S0169207019301918},
	doi = {10.1016/j.ijforecast.2019.05.014},
	pages = {428--441},
	number = {2},
	journaltitle = {International Journal of Forecasting},
	shortjournal = {International Journal of Forecasting},
	author = {Taylor, James W.},
	urldate = {2026-03-18},
	date = {2020-04},
	langid = {english},
	note = {{JCR分区}: Q1
影响因子: 7.1
5年影响因子: 7.5
中科院升级版小类分区: 经济学2区/管理学3区。
{SSCI}: Q1
{AJG}: 3
{ABDC}: A},
}

@misc{themathworksinc.2026,
	location = {Natick, Massachusetts, United States},
	title = {Deep learning toolbox: {TrainAutoencoder} (R2025b)},
	url = {https://au.mathworks.com/help/deeplearning/ref/trainautoencoder.html},
	publisher = {The {MathWorks} Inc.},
	author = {{The MathWorks, Inc.}},
	date = {2026},
}

@report{baselcommitteeonbankingsupervision2019,
	title = {Minimum capital requirements for market risk},
	url = {https://www.bis.org/bcbs/publ/d457.htm},
	institution = {Bank for International Settlements},
	author = {{Basel Committee on Banking Supervision}},
	date = {2019-01},
}

@article{olshausen1997a,
	title = {Sparse coding with an overcomplete basis set: A strategy employed by V1?},
	volume = {37},
	issn = {00426989},
	url = {https://linkinghub.elsevier.com/retrieve/pii/S0042698997001697},
	doi = {10.1016/S0042-6989(97)00169-7},
	shorttitle = {Sparse coding with an overcomplete basis set},
	pages = {3311--3325},
	number = {23},
	journaltitle = {Vision Research},
	shortjournal = {Vision Research},
	author = {Olshausen, Bruno A. and Field, David J.},
	urldate = {2026-03-18},
	date = {1997-12},
	note = {{JCR分区}: Q3
影响因子: 1.4
5年影响因子: 1.7
中科院升级版小类分区: 神经科学4区/眼科学4区/心理学4区。},
}

@article{hansen2016a,
	title = {Exponential {GARCH} Modeling With Realized Measures of Volatility},
	volume = {34},
	issn = {0735-0015},
	url = {https://www.jstor.org/stable/44166579},
	pages = {269--287},
	number = {2},
	journaltitle = {Journal of Business \& Economic Statistics},
	shortjournal = {J. Bus. Econom. Statist.},
	publisher = {[American Statistical Association, Taylor \& Francis, Ltd.]},
	author = {Hansen, Peter Reinhard and Huang, Zhuo},
	urldate = {2024-07-17},
	date = {2016},
	note = {{JCR分区}: Q1
影响因子: 2.5
{SSCI}: Q1
{ABDC}: A*
5年影响因子: 5.1
中科院升级版小类分区: 统计学与概率论2区/经济学3区/社会科学：数理方法3区。
{AJG}: 4},
}

@article{hansen2012,
	title = {Realized {GARCH}: a joint model for returns and realized measures of volatility},
	volume = {27},
	rights = {http://onlinelibrary.wiley.com/{termsAndConditions}\#vor},
	issn = {0883-7252, 1099-1255},
	url = {https://onlinelibrary.wiley.com/doi/10.1002/jae.1234},
	doi = {10.1002/jae.1234},
	shorttitle = {Realized {GARCH}},
	pages = {877--906},
	number = {6},
	journaltitle = {Journal of Applied Econometrics},
	shortjournal = {J. Appl. Econometrics},
	author = {Hansen, Peter Reinhard and Huang, Zhuo and Shek, Howard Howan},
	urldate = {2024-10-08},
	date = {2012-09},
	langid = {english},
	note = {{JCR分区}: Q1
影响因子: 3.1
{SSCI}: Q1
{ABDC}: A*
5年影响因子: 2.9
中科院升级版小类分区: 经济学3区/社会科学：数理方法3区。
{AJG}: 3},
}

@article{bollerslev1986,
	title = {Generalized autoregressive conditional heteroskedasticity},
	volume = {31},
	issn = {03044076},
	url = {https://linkinghub.elsevier.com/retrieve/pii/0304407686900631},
	doi = {10.1016/0304-4076(86)90063-1},
	pages = {307--327},
	number = {3},
	journaltitle = {Journal of Econometrics},
	shortjournal = {J. Econometrics},
	author = {Bollerslev, Tim},
	urldate = {2024-07-16},
	date = {1986-04},
	note = {{JCR分区}: Q1
影响因子: 4.0
{SSCI}: Q1
{ABDC}: A*
5年影响因子: 7.8
中科院升级版小类分区: 数学跨学科应用1区/社会科学：数理方法1区/经济学2区。
{AJG}: 4},
}

@article{barndorff-nielsen2009,
	title = {Realized kernels in practice: trades and quotes},
	volume = {12},
	issn = {1368-4221},
	url = {https://www.jstor.org/stable/23116045},
	shorttitle = {Realized kernels in practice},
	pages = {C1--C32},
	number = {3},
	journaltitle = {The Econometrics Journal},
	publisher = {[Royal Economic Society, Wiley]},
	author = {Barndorff-Nielsen, O. E. and Hansen, P. Reinhard and Lunde, A. and Shephard, N.},
	urldate = {2024-07-16},
	date = {2009},
	note = {{JCR分区}: Q1
影响因子: 7.0
{SSCI}: Q1
5年影响因子: 4.6
中科院升级版小类分区: 数学跨学科应用3区/社会科学：数理方法3区/统计学与概率论3区/经济学4区。
{AJG}: 3},
}

@article{barndorff-nielsen2008b,
	title = {Measuring Downside Risk - Realised Semivariance},
	issn = {1556-5068},
	url = {http://www.ssrn.com/abstract=1262194},
	doi = {10.2139/ssrn.1262194},
	journaltitle = {{SSRN} Electronic Journal},
	shortjournal = {{SSRN} Journal},
	author = {Barndorff-Nielsen, Ole E. and Kinnebrock, Silja and Shephard, Neil},
	urldate = {2024-10-18},
	date = {2008},
	langid = {english},
}

@article{baker2016,
	title = {Measuring Economic Policy Uncertainty*},
	volume = {131},
	issn = {0033-5533},
	url = {https://doi.org/10.1093/qje/qjw024},
	doi = {10.1093/qje/qjw024},
	pages = {1593--1636},
	number = {4},
	journaltitle = {The Quarterly Journal of Economics},
	shortjournal = {The Quarterly Journal of Economics},
	author = {Baker, Scott R. and Bloom, Nicholas and Davis, Steven J.},
	urldate = {2024-10-21},
	date = {2016-11-01},
}

@article{barndorff-nielsen2004,
	title = {Power and Bipower Variation with Stochastic Volatility and Jumps},
	volume = {2},
	issn = {1479-8409, 1479-8417},
	url = {https://academic.oup.com/jfec/article-lookup/doi/10.1093/jjfinec/nbh001},
	doi = {10.1093/jjfinec/nbh001},
	pages = {1--37},
	number = {1},
	journaltitle = {Journal of Financial Econometrics},
	shortjournal = {Journal of Financial Econometrics},
	author = {Barndorff-Nielsen, O. E.},
	urldate = {2024-10-17},
	date = {2004-12-01},
	langid = {english},
	note = {{JCR分区}: Q2
影响因子: 2.2
{SSCI}: Q2
{ABDC}: A*},
}

@article{andersen2003,
	title = {Modeling and Forecasting Realized Volatility},
	volume = {71},
	rights = {The Econometric Society 2003},
	issn = {1468-0262},
	url = {https://onlinelibrary.wiley.com/doi/abs/10.1111/1468-0262.00418},
	doi = {10.1111/1468-0262.00418},
	pages = {579--625},
	number = {2},
	journaltitle = {Econometrica},
	author = {Andersen, Torben G. and Bollerslev, Tim and Diebold, Francis X. and Labys, Paul},
	urldate = {2024-10-17},
	date = {2003},
	langid = {english},
	note = {\_eprint: https://onlinelibrary.wiley.com/doi/pdf/10.1111/1468-0262.00418},
}

@article{zhang2005,
	title = {A Tale of Two Time Scales: Determining Integrated Volatility With Noisy High-Frequency Data},
	volume = {100},
	issn = {0162-1459, 1537-274X},
	url = {http://www.tandfonline.com/doi/abs/10.1198/016214505000000169},
	doi = {10.1198/016214505000000169},
	pages = {1394--1411},
	number = {472},
	journaltitle = {Journal of the American Statistical Association},
	shortjournal = {J. Am. Stat. Assoc.},
	publisher = {{ASA} Website},
	author = {Zhang, Lan and Mykland, Per A and Aït-Sahalia, Yacine},
	urldate = {2024-11-04},
	date = {2005-12},
	note = {\_eprint: https://doi.org/10.1198/016214505000000169
{JCR分区}: Q1
影响因子: 3.0
{ABDC}: A*
5年影响因子: 4.8
中科院升级版Top分区: 数学{TOP}
中科院升级版小类分区: 统计学与概率论2区。
{AJG}: 4},
}

@article{andersen1998,
	title = {Answering the Skeptics: Yes, Standard Volatility Models do Provide Accurate Forecasts},
	volume = {39},
	issn = {0020-6598},
	url = {https://www.jstor.org/stable/2527343},
	doi = {10.2307/2527343},
	shorttitle = {Answering the Skeptics},
	pages = {885--905},
	number = {4},
	journaltitle = {International Economic Review},
	shortjournal = {Int. Econom. Rev.},
	publisher = {[Economics Department of the University of Pennsylvania, Wiley, Institute of Social and Economic Research, Osaka University]},
	author = {Andersen, Torben G. and Bollerslev, Tim},
	urldate = {2024-10-13},
	date = {1998},
	note = {{JCR分区}: Q3
影响因子: 1.3
{SSCI}: Q3
{ABDC}: A*},
}

@article{barndorff-nielsen2008,
	title = {Designing Realised Kernels to Measure the Ex-Post Variation of Equity Prices in the Presence of Noise},
	issn = {1556-5068},
	url = {http://www.ssrn.com/abstract=620203},
	doi = {10.2139/ssrn.620203},
	journaltitle = {{SSRN} Electronic Journal},
	shortjournal = {{SSRN} Journal},
	author = {Barndorff-Nielsen, Ole E. and Hansen, Peter Reinhard and Lunde, Asger and Shephard, Neil},
	urldate = {2024-10-17},
	date = {2008},
	langid = {english},
}

@article{ikeda2015,
	title = {Two-Scale Realized Kernels: A Univariate Case},
	volume = {13},
	issn = {1479-8409, 1479-8417},
	url = {https://academic.oup.com/jfec/article-lookup/doi/10.1093/jjfinec/nbt024},
	doi = {10.1093/jjfinec/nbt024},
	shorttitle = {Two-Scale Realized Kernels},
	pages = {126--165},
	number = {1},
	journaltitle = {Journal of Financial Econometrics},
	shortjournal = {Journal of Financial Econometrics},
	author = {Ikeda, S. S.},
	urldate = {2024-10-18},
	date = {2015-01-01},
	langid = {english},
	note = {{JCR分区}: Q2
影响因子: 2.2
5年影响因子: 2.9
中科院升级版小类分区: 商业：财政与金融4区/经济学4区。
{SSCI}: Q2
{AJG}: 3
{ABDC}: A*},
}

@article{andersen2012,
	title = {Jump-robust volatility estimation using nearest neighbor truncation},
	volume = {169},
	issn = {03044076},
	url = {https://linkinghub.elsevier.com/retrieve/pii/S0304407612000127},
	doi = {10.1016/j.jeconom.2012.01.011},
	series = {Recent Advances in Panel Data, Nonlinear and Nonparametric Models: A Festschrift in Honor of Peter C.B. Phillips},
	pages = {75--93},
	number = {1},
	journaltitle = {Journal of Econometrics},
	shortjournal = {J. Econometrics},
	author = {Andersen, Torben G. and Dobrev, Dobrislav and Schaumburg, Ernst},
	urldate = {2024-10-18},
	date = {2012-07},
	note = {{JCR分区}: Q1
影响因子: 4.0
{SSCI}: Q1
{ABDC}: A*
5年影响因子: 7.8
中科院升级版小类分区: 数学跨学科应用1区/社会科学：数理方法1区/经济学2区。
{AJG}: 4},
}

@article{gneiting2011,
	title = {Making and evaluating point forecasts},
	volume = {106},
	issn = {0162-1459, 1537-274X},
	url = {http://www.tandfonline.com/doi/abs/10.1198/jasa.2011.r10138},
	doi = {10.1198/jasa.2011.r10138},
	pages = {746--762},
	number = {494},
	journaltitle = {Journal of the American Statistical Association},
	shortjournal = {Journal of the American Statistical Association},
	publisher = {Taylor \& Francis},
	author = {Gneiting, Tilmann},
	urldate = {2026-03-24},
	date = {2011-06},
	note = {\_eprint: https://doi.org/10.1198/jasa.2011.r10138
{JCR分区}: Q1
影响因子: 3.0
5年影响因子: 4.8
中科院升级版Top分区: 数学{TOP}
中科院升级版小类分区: 统计学与概率论2区。
{AJG}: 4
{ABDC}: A*},
}

@article{fissler2016,
	title = {Higher order elicitability and osband’s principle},
	volume = {44},
	issn = {0090-5364},
	url = {https://projecteuclid.org/journals/annals-of-statistics/volume-44/issue-4/Higher-order-elicitability-and-Osbands-principle/10.1214/16-AOS1439.full},
	doi = {10.1214/16-AOS1439},
	pages = {1680--1707},
	number = {4},
	journaltitle = {The Annals of Statistics},
	shortjournal = {Ann. Statist.},
	publisher = {Institute of Mathematical Statistics},
	author = {Fissler, Tobias and Ziegel, Johanna F.},
	urldate = {2026-03-25},
	date = {2016-08-01},
	langid = {english},
	note = {{JCR分区}: Q1
影响因子: 3.7
5年影响因子: 5.9
中科院升级版Top分区: 数学{TOP}
中科院升级版小类分区: 统计学与概率论1区。
{AJG}: 4*},
}

@article{hinton2006,
	title = {Reducing the Dimensionality of Data with Neural Networks},
	volume = {313},
	issn = {0036-8075, 1095-9203},
	url = {https://www.science.org/doi/10.1126/science.1127647},
	doi = {10.1126/science.1127647},
	pages = {504--507},
	number = {5786},
	journaltitle = {Science},
	shortjournal = {Science},
	author = {Hinton, G. E. and Salakhutdinov, R. R.},
	urldate = {2024-07-17},
	date = {2006-07-28},
	langid = {english},
	note = {{JCR分区}: Q1
影响因子: 45.8
5年影响因子: 49.7
中科院升级版Top分区: 综合性期刊{TOP}
中科院升级版小类分区: 综合性期刊1区。},
}

@article{ghysels2007,
	title = {{MIDAS} regressions: Further results and new directions},
	volume = {26},
	issn = {0747-4938, 1532-4168},
	url = {http://www.tandfonline.com/doi/abs/10.1080/07474930600972467},
	doi = {10.1080/07474930600972467},
	shorttitle = {{MIDAS} Regressions},
	pages = {53--90},
	number = {1},
	journaltitle = {Econometric Reviews},
	shortjournal = {Econom. Rev.},
	publisher = {Taylor \& Francis},
	author = {Ghysels, Eric and Sinko, Arthur and Valkanov, Rossen},
	urldate = {2026-03-03},
	date = {2007-02-05},
	note = {\_eprint: https://doi.org/10.1080/07474930600972467
{JCR分区}: Q3
影响因子: 1.0
5年影响因子: 1.6
中科院升级版小类分区: 经济学4区/数学跨学科应用4区/社会科学：数理方法4区/统计学与概率论4区。
{SSCI}: Q3
{ABDC}: A
{AJG}: 3},
}
